\documentclass{dissertation}

\makeatletter
\@ifundefined{widering}{}{%
	
}
\makeatother

\AtBeginDocument{%
	\DeclareSymbolFont{operators}{OT1}{ntxtlf}{m}{n}
	\SetSymbolFont{operators}{bold}{OT1}{ntxtlf}{b}{n}
}
\usepackage{bbm}
\newcommand{\1}{\mathbbm{1}}
\usepackage{bm}

\usepackage{setspace}
\makeglossaries

\usepackage[skip=4pt, indent=16pt]{parskip}

\usepackage{lipsum}

\begin{document}

\title{Enhanced Digitized Adiabatic Quantum Factorization Algorithm Using Null-Space Encoding}

\frontmatter

\begin{titlepage}
\newgeometry{top=3cm,bottom=3cm,left=2.5cm,right=2.5cm} 

\begin{center}

\noindent
\begin{minipage}[t]{0.45\textwidth}
    \raggedright
    \includegraphics[height=1cm]{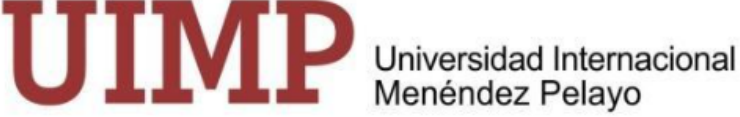}
\end{minipage}%
\hfill
\begin{minipage}[t]{0.45\textwidth}
    \raggedleft
    \includegraphics[height=1.1cm]{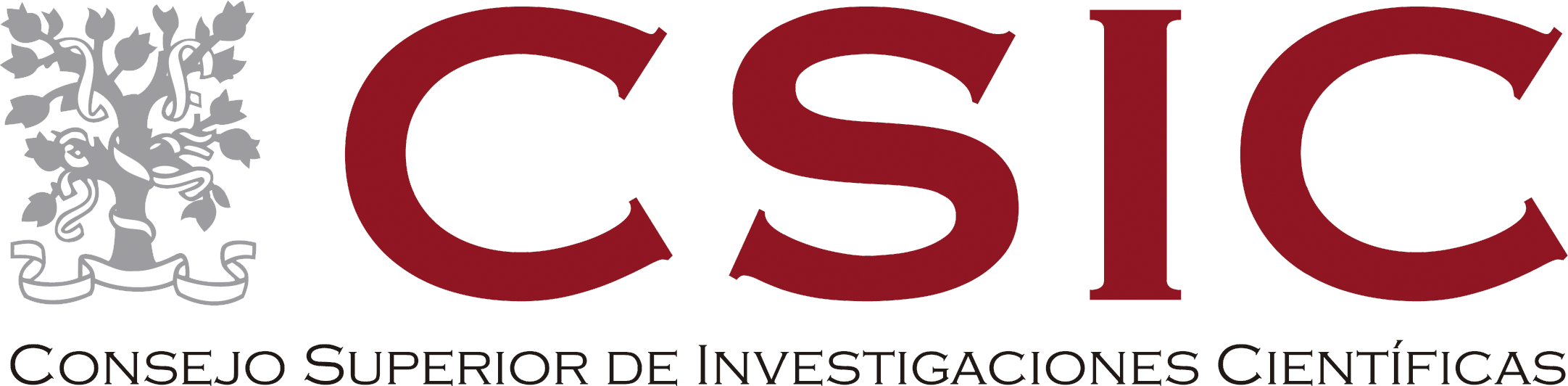}
\end{minipage}

\vspace{4cm}

\includegraphics[width=0.95\textwidth]{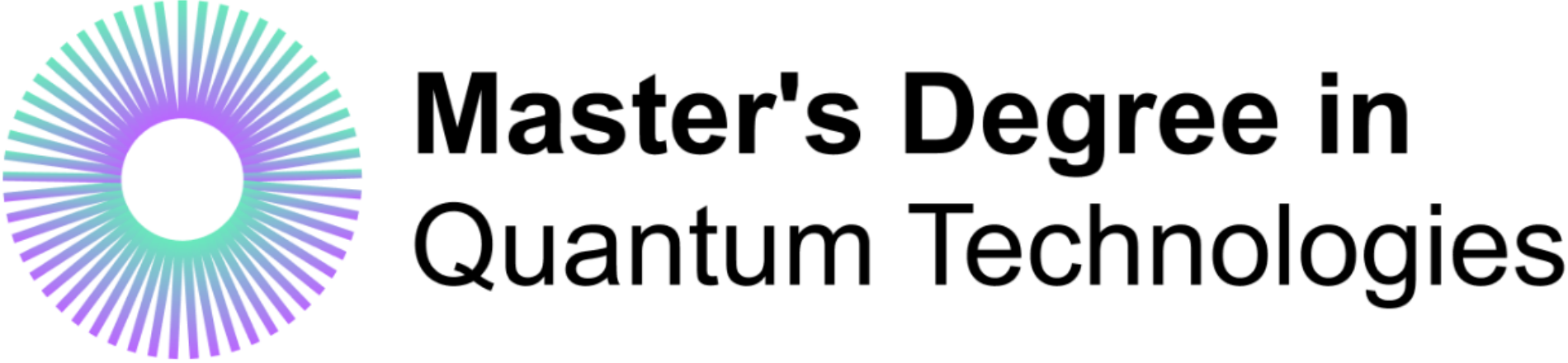}

\vspace{1cm}
{\Large Academic Year: 2024 -- 2025 \par}

\vspace{1cm}
{\huge
Enhanced Digitized Adiabatic Quantum\\[0.2cm]
Factorization Algorithm Using Null-Space Encoding\par}

\vspace{5cm}

{\Large Author: Felip Pellicer Benedicto \par}
\vspace{0.3cm}
{\Large Directors: Alan Costa dos Santos, Juan José García-Ripoll \par}

\end{center}

\restoregeometry
\end{titlepage}


\thispagestyle{empty}
\tableofcontents
\let\cleardoublepage\clearpage

\chapter*{Abstract}
\addcontentsline{toc}{chapter}{Abstract}
\setheader{Abstract}

Integer factorization is a computational problem of fundamental importance in cybersecurity and secure communications, as its difficulty form the basis of modern public-key cryptography. While Shor's algorithm can solve this problem efficiently on a universal quantum computer, near-term devices require alternative approaches. The Adiabatic Factorization Algorithm and its digitized counterparts offer a promising NISQ-era pathway but suffer from high-order many-body interactions that are difficult to implement. In this work, we propose a modified QAOA-based factorization protocol that simplifies the interacting Hamiltonian to include only two-body terms, significantly reducing its experimental complexity. Numerical simulations show that this method achieves comparable or higher fidelities than the standard protocol, while requiring fewer quantum resources and converging more rapidly for problem instances up to eight qubits. We analyze the characteristic fidelity behavior introduced by the Hamiltonian modification. Additionally, we report on simulations with alternative cost-function definitions that frequently yielded improved performance.

\chapter*{Resumen}
\addcontentsline{toc}{chapter}{Resumen}
\setheader{Resumen}

\noindent La factorización de números enteros es un problema computacional de gran relevancia en los campos de la ciberseguridad y las comunicaciones seguras, ya que su dificultad sirve como base para la criptografía de clave pública moderna. Mientras que el algoritmo de Shor resuelve este problema de forma eficiente en un ordenador cuántico universal, los dispositivos a corto plazo requieren de enfoques diferentes. El Algoritmo de Factorización Adiabática y sus versiones digitalizadas ofrecen un prometedor camino a seguir en la era NISQ (por sus siglas en inglés \textit{Noisy Intermediate-Scale Quantum}), pero sufren de interacciones de muchos cuerpos que son difíciles de implementar. En este trabajo, proponemos un protocolo de factorización basado en QAOA modificado que simplifica el Hamiltoniano de interacción para incluir únicamente interacciones de dos cuerpos, reduciendo significativamente su complejidad experimental. Simulaciones numéricas demuestran que este método logra fidelidades comparablas o mayores a las obtenidas con el protocolo estándar, a la vez que requiere un menor número de recursos cuánticos y convergiendo más rápidamente para problemas de hasta ocho qubits. Analizamos el comportamiento característico en la evolución de la fidelidad introducido por la modificación en el Hamiltoniano. Además, utilizamos definiciones alternativas en la función de coste que normalmente conllevaron mejoras de rendimiento.

\mainmatter

\thumbtrue

\chapter{Introduction}
\label{introduction}

\dropcap{Q}uantum computing has emerged as the most promising paradigm for solving problems that are classically intractable. A central example is the integer factorization problem, which consists of decomposing a semiprime number $N$ into its prime constituents. While seemingly simple, no known classical algorithm can factorize large integers efficiently, and the best classical methods --- such as the general number field sieve --- scale superpolynomially with input size~\cite{montgomery_cwi_1994}. This computational asymmetry forms the foundation of widely used cryptographic protocols, most notably RSA (Rivest-Shamir-Adleman), whose security relies on the practical difficulty of factorizing large semiprimes~\cite{rivest_method_1978}.

One of the most celebrated achievements in quantum computing was the discovery of Shor's algorithm~\cite{shor_algorithms_1994}, which demonstrated that integer factorization can be solved exponentially faster on a quantum computer than with the best-known classical algorithm, thereby posing a direct threat to RSA-based cryptography. However, the full realization of this algorithm is currently challenging due to quantum hardware limitations in Noisy Intermediate-Scale Quantum (NISQ) devices.

The constraints imposed by the NISQ era have motivated alternative methods to attempt to solve the same problems by using fewer quantum resources. One such approach is the Adiabatic Quantum Computation (AQC)~\cite{albash_adiabatic_2018}, which takes advantage of the robustness of adiabatic evolution to bring the system from a known state to a final state that encodes the solution of a problem. Another approach is the Variational Quantum Algorithm (VQA)~\cite{cerezo_variational_2021}, which is a hybrid algorithm that uses quantum hardware for state evolution, and classical routines for optimization purposes.

In this work, we propose and analyze a version of the Digitized Adiabatic Quantum Factorization~\cite{hegade_digitized_2021} algorithm that incorporates a variation in the cost Hamiltonian. By this, we aim to provide a more resource-efficient and scalable method for integer factorization.

\let\clearpage

\section{Circuit Model of Quantum Computation}
\label{Section:GateModelQC}

The circuit or gate-based model of quantum computation is the most widely studied framework and the foundation of many quantum algorithms, including Shor's and Grover's algorithms~\cite{nielsen00}. In this model, computation proceeds through the application of a sequence of unitary gates, which evolve the state of the qubits in a discrete, step-wise fashion.

Mathematically, a quantum circuit implements a unitary transformation $U$ on the initial quantum state, typically chosen as $\ket{0}^{\otimes n}$. This transformation is decomposed into a series of quantum gates from a universal set of gates. Measurement in the computational basis is performed at the end of the circuit to extract classical information from the quantum circuit.

Universality is a crucial property of this model, as it ensures that any unitary transformation --- and therefore any quantum algorithm --- can be implemented using only a finite set of gates. This makes the circuit model a general-purpose framework, capable of simulating any other model of quantum computation.

Gate-based quantum computation aligns well with digital control paradigms, and most existing quantum hardware platforms, such as superconducting qubits and trapped ions, are designed to implement this model. However, the depth and width of circuits that can be reliably executed on near-term devices are limited by noise, decoherence, and imperfect gate fidelities.
\section{Adiabatic Quantum Computation}
\label{Section:AQC}
The adiabatic theorem of quantum mechanics provides the foundation for an alternative model of quantum computation based on continuous-time evolution. Consider a time-dependent Hamiltonian $H(t)$ with a discrete and non-degenerate spectrum. Thus we can define its instantaneous eigenstates and eigenenergies by
\begin{equation}
    \hat{H}(t) \ket{n(t)} = E_n (t) \ket{n(t)}.
    \label{eq:instantaneous_eigenstates}
\end{equation}

\begin{figure}[h]
    \centering
    \includegraphics[width=0.75\textwidth]{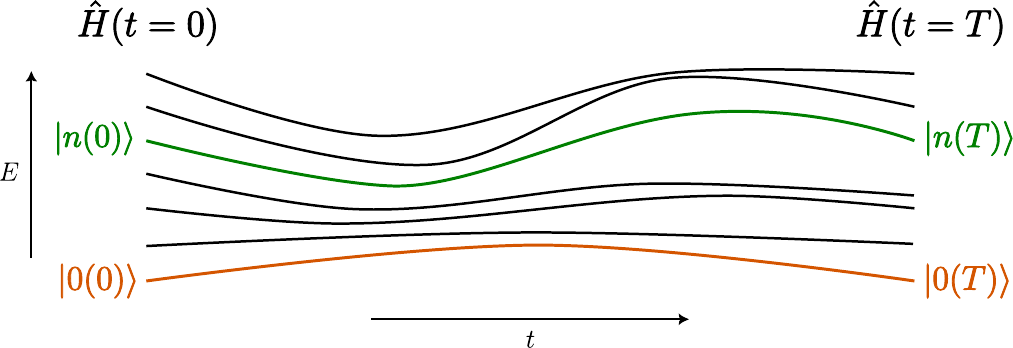}
    \caption{Schematic of an adiabatic passage. The orange line represents the Hamiltonian's ground state, while the green a Hamiltonian's eigenstate different from the ground state.}
    \label{fig:adiabatic_passage}
\end{figure}

\noindent As depicted in Fig.~\ref{fig:adiabatic_passage}, the adiabatic theorem states that a quantum system initially prepared in an eigenstate $\ket{n(0)}$ of $H(0)$ will remain in the corresponding instantaneous eigenstate $\ket{n(t)}$ throughout the evolution, provided the spectrum remains gapped and the Hamiltonian changes sufficiently slowly~\cite{sarandy_consistency_2004}. This condition is generally formulated as
\begin{equation}
    T \gg \frac{\mathcal{F}}{\mathcal{G}^2}, 
    \qquad 
    \mathcal{F} = \max_{0 \leq s \leq 1} \big| \bra{k}\,\tfrac{d}{ds}\hat{H}(s)\,\ket{n} \big|, 
    \qquad
    \mathcal{G} = \min_{0 \leq s \leq 1} |g_{nk}(s)| ,
    \label{eq:adiabatic_condition}
\end{equation}
where $T$ is the total runtime, $g_{nk}(s) = E_n(s) - E_k(s)$ is the instantaneous energy gap between levels $n$ and $k$, and $s=t/T$ is the normalized time. In practice, satisfying the adiabatic condition typically requires very long evolution times, which makes implementations vulnerable to decoherence and other noise sources.

Despite the adiabatic theorem works for any eigenstate, in practice it is customary to focus on ground state adiabatic passages. This is because ground states are typically more robust against decoherence and thermal excitations, whenever they are energetically isolated from higher-energy levels. Moreover, for many computational problems, particularly those related to optimization problems, it is possible to construct a Hamiltonian $\hat{H}_\mathrm{P}$ whose ground state encodes the solution to the instance under consideration.

In the adiabatic quantum computation model, the Hamiltonian is interpolated between an initial Hamiltonian $\hat{H}_0$, with a known and easily preparable ground state, and the problem Hamiltonian $\hat{H}_\mathrm{P}$, according to a schedule~\cite{albash_adiabatic_2018}:
\begin{equation}
    \hat{H}(t) = \big[1 - f(t)\big] \hat{H}_0 + f(t) \hat{H}_\mathrm{P}, \quad f(t) \in [0,1],
    \label{eq:adiabatic_passage}
\end{equation}
where $f(t)$ is a smooth, monotonic function such that $f(0)=0$ and $f(T)=1$. The problem Hamiltonian $\hat{H}_\mathrm{P}$ is designed so that its ground state corresponds to the solution of the computational task.

Aharonov et al.~\cite{aharonov_adiabatic_2004} proved that this model is computationally equivalent to the standard circuit model, meaning that any adiabatic process can, in principle, be reproduced digitally with comparable efficiency. This equivalence opens the door to a variety of gate-based approaches that draw inspiration from adiabatic evolution while remaining fully compatible with the circuit model of quantum computation.

\subsection{Adiabatic Quantum Annealers}
An important and widely studied application of adiabatic evolution in quantum devices is Adiabatic Quantum Annealing (AQA), where the goal is to find low-energy configurations of a classical cost function mapped onto a quantum Hamiltonian. AQA is most often applied to Quadratic Unconstrained Binary Optimization (QUBO) problems~\cite{kadowaki_quantum_1998}, which can be exactly reformulated as a 2-body interaction Ising Hamiltonian of the form~\cite{albash_adiabatic_2018}
\begin{equation}
    \hat{H}_\mathrm{P} = \sum_i h_i \hat{\sigma}_i^z + \sum_{i<j} J_{ij} \hat{\sigma}_i^z \hat{\sigma}_j^z\,,
    \label{eq:ising_hamiltonian}
\end{equation}
where $h_i$ are local fields, $J_{ij}$ represent coupling strengths between qubits, and $\hat{\sigma}_i^z$ are Pauli-Z operators acting on the $i$-th qubit, with eigenstates $\ket{0}$ and $\ket{1}$ satisfying $\hat{\sigma}^z \ket{0} = \ket{0}$ and $\hat{\sigma}^z \ket{1} = -\ket{1}$. The task is to drive the system from an initial, easily preparable ground state towards the ground state of $H_\mathrm{P}$, which encodes the optimal solution to the problem at hand.

In a typical AQA protocol, the evolution begins with an initial Hamiltonian of a system with a X-oriented field component of the form
\begin{equation}
    \hat{H}_0 = - \sum_i \hat{\sigma}_i^x\,.
    \label{eq:transverse_field_hamiltonian}
\end{equation}
where $\hat{\sigma}^x = \ket{0}\bra{1}+\ket{1}\bra{0}$ and whose ground state is straightforward to prepare. The system is then driven according to the interpolation scheme in Eq.~\eqref{eq:adiabatic_passage}. The aim is to adiabatically steer the system from the easily preparable ground state $H_0$ to the ground state of $H_\mathrm{P}$, which encodes the solution to the problem of interest.

This procedure is feasible for problems that can be expressed in the QUBO form, as they can be mapped to an Ising Hamiltonian with only two-body interactions. This approach has been physically implemented in superconducting-qubit devices, most notably in the commercial quantum annealers developed by D-Wave. However, as we shall see in chapter~\ref{Chapter:Factorization}, the problem of interest in this thesis, integer factorization, does not directly fall into the QUBO class, as its Hamiltonian contains three- and four-body interaction terms.
\section{Quantum Approximate Optimization Algorithm}
\label{Section:QAOA}

As mentioned previously, AQC is naturally formulated in platforms where fields and controls evolve continuously in time. Implementing this paradigm directly on gate-based quantum devices, however, can be challenging, since these platforms rely on discrete operations. A way to bridge this gap is provided by the Quantum Approximate Optimization Algorithm (QAOA), a hybrid quantum-classical method introduced by Farhi et al.~\cite{farhi_quantum_2014} to tackle combinatorial optimization problems. The algorithm aims to approximate the ground state of a cost Hamiltonian whose minimum encodes the optimal solution. Thanks to its shallow circuit depth and variational structure, which delegates part of the computational workload to a classical optimizer, QAOA is particularly well suited to near-term quantum devices.

At its core, QAOA constructs a parametrized quantum state (ansatz) by sequentially applying two alternating types of unitaries derived from two Hamiltonians:
\begin{itemize}
    \item The cost Hamiltonian $H_C$, which encodes the objective function of the optimization
    problem. This is typically written in the form of an Ising Hamiltonian.
    \item The mixing Hamiltonian $H_M$, which introduces transitions between computational
    basis states to enable exploration of the solution space. A common choice, justified by Eq.~\eqref{eq:transverse_field_hamiltonian}, is:
    \begin{equation}
        \hat{H}_M = - \sum_i \hat{\sigma}_i^x,
        \label{eq:mixing_hamiltonian}
    \end{equation}
    where $\hat{\sigma}_i^x$ is the Pauli-X operator acting on qubit $i$.
\end{itemize}

The algorithm starts with the initial state $\ket{\psi_0} = \ket{+}^{\otimes n}$, which is the ground state of $H_M$ and a uniform superposition over all computational basis states, where $\ket{+} = \dfrac{1}{\sqrt{2}} \big(\ket{0} + \ket{1}\big)$. The QAOA ansatz with $p$ layers is constructed as:
\begin{equation}
    \ket{\psi_p(\bm{\gamma}, \bm{\beta})} = e^{-i \beta_p \hat{H}_M} e^{- i \gamma_p \hat{H}_C} \cdots
    e^{-i \beta_1 \hat{H}_M} e^{- i \gamma_1 \hat{H}_C} \ket{+}^{\otimes n},
    \label{eq:qaoa_state_evolution}
\end{equation}
where $\bm{\gamma} = (\gamma_1, \dots, \gamma_p)$ and $\bm{\beta} = (\beta_1, \dots, \beta_p)$ are real variational parameters to be optimized.

The performance of a QAOA instance is assessed by evaluating the expectation value of the cost Hamiltonian in the ansatz state:
\begin{equation}
    F_p (\bm{\gamma}, \bm{\beta}) = \bra{\psi_p (\bm{\gamma}, \bm{\beta})} \hat{H}_C \ket{\psi_p (\bm{\gamma}, \bm{\beta})}.
    \label{eq:cost_function}
\end{equation}
This expectation value serves as the cost function for the classical optimizer. The parameters $(\bm{\gamma}, \bm{\beta})$ are iteratively updated to minimize $F_p$, with quantum circuits being re-evaluated at each step until convergence.

\begin{figure}[h]
    \centering
    \includegraphics[width=0.6\textwidth]{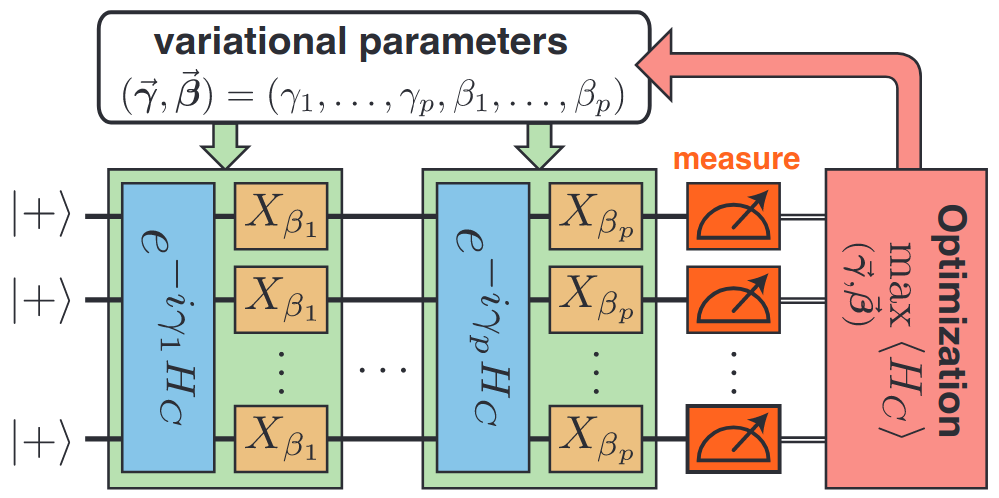}
    \caption{Diagram of a $p$-layer QAOA circuit. Starting from the initial state $\ket{+}^{\otimes n}$,the circuit alternates between applying the unitaries $e^{-i \gamma_i \hat{H}_C}$ and $e^{-i \beta_i \hat{H}_M}$ for $i = 1$ to $p$. The final state is measured to estimate the expectation value $\langle \hat{H}_C \rangle$, which is passed to a classical optimizer. This process is repeated until convergence.}
    \vspace{0.3em}
    \small\textit{Source: Adapted from Zhou et al., 2019.~\cite{zhou_quantum_2020}}
    \label{fig:qaoa}
\end{figure}

The Quantum Approximate Optimization Algorithm stands out as one of the most promising approaches for near-term quantum advantage. From a complexity-theoretic perspective, Farhi et al.~\cite{farhi_quantum_2019} argue that sampling from the output distribution of even the shallowest version of QAOA ($p=1$) may already be classically intractable, highlighting its potential as an early demonstration of quantum supremacy.

Beyond this theoretical motivation, QAOA is attractive because of its relatively simple structure and hardware efficiency. Ho and Hsieh~\cite{ho_efficient_2019} introduce a related variational protocol (VQCS), showing that alternating unitaries generated by simple, local Hamiltonians can efficiently prepare non-trivial quantum states and can be implemented on current platforms such as trapped ions and superconducting qubits.

Additionally, QAOA's modularity and variational flexibility make it adaptable to a wide range of applications, including factorization. In particular, Anschuetz et al.~\cite{anschuetz_variational_2018} demonstrate a QAOA-inspired approach to integer factorization, showing that hybrid quantum-classical heuristics can extend QAOA's reach to classically hard problems even on today's noisy devices.

The Quantum Approximate Optimization Algorithm has been extensively investigated since its introduction, leading to a broad landscape of theoretical analyses, practical  implementations, and performance-enhancement techniques. Numerous variants have been proposed, ranging from modified ansatze to parameter initialization heuristics and adaptive layer-scaling procedures. This diversity reflects both the flexibility of the QAOA framework and the complexity of identifying implementations that perform well across different problem instances. This richness also underscores the importance of carefully specifying the algorithmic configuration adopted in any particular study, so that performance comparisons can be made in a meaningful way.

\chapter{Digitized Factorization}
\label{Chapter:Factorization}

\dropcap{T}his chapter explores two quantum approaches to the factorization problem. First, we review
Shor's algorithm and its quantum Fourier transform-based structure. Next, we introduce an alternative method for factorization based on adiabatic quantum computation, referred to as \textit{adiabatic factorization}~\cite{peng_quantum_2008}, which may serve as a potentially viable approach for factorization on analog quantum computers. We briefly discuss the limitations of this method, primarily arising from the implementation of many-qubit gates, which pose significant challenges in practical experimental settings. To address this issue, we present our solution in the form of a \textit{linearized adiabatic} Hamiltonian for factorization. Finally, we conclude this section with a discussion of the methods employed to implement each algorithm, setting the stage for subsequent comparisons and analysis.
\section{Shor's Factorization Algorithm}

In computer science, an algorithm is considered to be efficient when the number of steps
of the algorithm grows as a polynomial in the input size. For the problem of integer factorization,
the input is a semiprime number $N$, and the input size is measured as the number of bits
required to represent it, i.e., $\log N$. The best known classical algorithm for factorizing
scales superpolynomially with input size, making the problem computationally hard for large $N$.
\begin{equation}
    \mathcal{O} \bigg( \exp \big( c(\log N)^{1/3} (\log \log N)^{2/3} \big) \bigg).
    \label{eq:field_sieve_scaling}
\end{equation}

In 1994, Peter Shor introduced a quantum algorithm that factorizes integers in polynomial time,
marking a landmark result in quantum computing. Shor's algorithm runs in time
\begin{equation}
    \mathcal{O} \big( (\log N)^3 \big),
    \label{eq:shor_scaling}
\end{equation}
dramatically outperforming the best classical method~\cite{nielsen00}. The algorithm relies on the quantum
circuit model, making it a gate-based approach to factorization and one of the strongest motivations
for the development of quantum computers.

The core idea of Shor's algorithm is to reduce factorization to the problem of order finding:
given a number $a$ coprime to $N$, find the smallest integer $r$ such that
\begin{equation}
    a^r = 1 \mod N.
    \label{eq:order_finding}
\end{equation}
Once the order $r$ is known, under certain conditions, one can recover a nontrivial factor of
$N$ using elementary number-theoretic arguments~\cite{nielsen00}.

The quantum part of the algorithm is used to efficiently find this order $r$ using
period-finding techniques. This is achieved by preparing a quantum superpositioin and applying the
quantum Fourier transform (QFT) to extract information about the period of the function
$f(x) = a^x \mod N$. The classical post-processing step then uses continued fractions to extract $r$
and attempt to derive the prime factors of $N$.

Despite its theoretical significance, implementing Shor's algorithm at scale requires a
large number of qubits, long coherence times, and fault-tolerant error correction,
which remain beyond the reach of current quantum hardware. As such, while Shor's algorithm remains
the most efficient known method for factorizing in the long-term quantum regime, alternative approaches
--- such as adiabatic and variational algorithms --- are being explored for use in the NISQ era.
\section{Adiabatic Factorization Algorithm}
\label{Section:AFA}

Before introducing the details of the adiabatic factorization algorithm, it is important to clarify its relationship to Shor’s algorithm. Although both approaches aim to solve the integer factorization problem, the adiabatic method presented here is not an adiabatic implementation of Shor’s algorithm. Instead, it represents a distinct strategy for factorization based on adiabatic evolution, which we refer to in this work as \textit{adiabatic factorization}.

The problem of integer factorization can be formulated as a constrained search over pairs of natural numbers $p$ and $q$ such that
\begin{equation}
	N = p \times q.
	\label{eq:integer_factorization}
\end{equation}
In the context of adiabatic quantum computation we aim to encode this constraint into the ground state of a problem Hamiltonian, allowing the solution to emerge through adiabatic evolution.

To encode candidate solutions $p$ and $q$, we adopt a binary representation of natural numbers. Any natural number $X$ can be expressed as:
\begin{equation}
	X = \sum_{j=0}^{n_\text{bits} - 1} 2^j x_j,
	\label{eq:binary_integer}
\end{equation}
where each $x_j \in \{0,1\}$ and the bit string $\mathbf{x} = x_{n_\text{bits}-1} \dots x_0$ represents the binary encoding of $X$. In their approach, Peng et al.~\cite{peng_quantum_2008} exploit the fact that the factors $p$ and $q$ of an odd composite number can be rewritten as:
\begin{equation}
	\begin{cases}
		p = 2p' + 1 \\
		q = 2q' + 1
	\end{cases} .
	\label{eq:factors_simplification}
\end{equation}
It can be proved that the $n_p$ and $n_q$ are an upper bound on the number of qubits required to represent $p'$ and $q'$, respectively:
\begin{equation}
	\begin{cases}
		n_p = m\,\big(\lfloor \sqrt{N} \rfloor_o\big) - 1 \\[2ex]
		n_q = m\bigg(\left\lfloor \dfrac{N}{3} \right\rfloor \bigg) - 1
	\end{cases} ,
	\label{eq:factors_num_bits}
\end{equation}
where $\lfloor a \rfloor \big(\lfloor a \rfloor_o\big)$ denotes the largest (odd) integer not larger than a, while $m(b)$ denotes the smallest number of bits required for representing $b$~\cite{peng_quantum_2008}. Then, the adiabatic factorization algorithm will make use of $n = n_p + n_q$ qubits. The full quantum state 
\begin{equation}
	\ket{\Psi} = \ket*{\Psi_{p'}} \otimes \ket*{\Psi_{q'}},
	\label{eq:full_quantum_state}
\end{equation}
where $\ket*{\Psi_{p'}} = \ket*{\psi_{1}}\otimes \cdots \otimes \ket*{\psi_{n_p}}$ and $\ket*{\Psi_{q'}} = \ket*{\psi_{n_p+1}}\otimes \cdots \otimes \ket*{\psi_{n_p +n_q}}$.

To encode the factorization constraint into the adiabatic quantum framework, we define the objective function as ~\cite{peng_quantum_2008}:
\begin{equation}
	f(p,q) = \big(N - p \times q \big)^2,
\end{equation}
such that its global minimum $f(p,q)=0$ corresponds to valid factors. Translating this into a Hamiltonian acting on the computational basis, we obtain the quadratic problem Hamiltonian:
\begin{equation}
	\hat{H}_\mathrm{QP} = \bigg[ N \1 - \bigg( \sum_{\ell=1}^{n_p} 2^\ell \hat{x}_\ell + \1 \bigg)
	\bigg( \sum_{m=1}^{n_q} 2^m \hat{y}_m + \1 \bigg) \bigg]^2\,,
	\label{eq:quadratic_problem_hamiltonian}
\end{equation}
where $\hat{x}_\ell = \dfrac{\1 - \hat{\sigma}_\ell^z}{2}$ and $\hat{y}_m = \dfrac{\1 - \hat{\sigma}_m^z}{2}$ are the number operators acting on the qubits encoding $p'$ and $q'$, respectively. The solution to the factorization problem is encoded in the ground state of $\hat{H}_\mathrm{QP}$.

The initial state of the system is prepared as:
\begin{equation}
	\ket{\psi(0)} = \ket{+}^{\otimes n}\,,
	\label{eq:initial_state}
\end{equation}
corresponding to the ground state of the initial Hamiltonian $\hat{H}_0$ defined in Eq.~\eqref{eq:transverse_field_hamiltonian}.
Following the adiabatic theorem, if the evolution from $\hat{H}_0$ to $\hat{H}_\mathrm{QP}$ is slow enough, the system will remain in the instantaneous ground state, reaching the ground state of $\hat{H}_\mathrm{QP}$ at the end of the protocol. This final state encodes the solution to the factorization problem.

Despite its conceptual clarity, the Hamiltonian $\hat{H}_\mathrm{QP}$ includes multiqubit interactions, such as three- and four-body terms of the form $\hat{\sigma}_\ell^z \hat{\sigma}_m^z \hat{\sigma}_k^z$ and $\hat{\sigma}_\ell^z \hat{\sigma}_m^z \hat{\sigma}_k^z \hat{\sigma}_n^z$. These many-body interactions are difficult to implement, since they require one to bring all involved qubits together and make them interact in a controlled way, resulting in prone-to-error processes. The need to avoid those terms is of pivotal interest to provide efficient quantum algorithms.

\section{Digitized Adiabatic Quantum Factorization}
Although the adiabatic model is often formulated in terms of continuous time evolution, it can be simulated efficiently using gate-based quantum computation through a process known as digitization.

In the digitized approach, the continuous adiabatic evolution governed by a time-dependent Hamiltonian as of Eq.~\eqref{eq:adiabatic_passage} is approximated by a sequence of quantum gates through trotterization, breaking the total evolution into small time slices. Each slice is implemented as a layer in a quantum circuit, simulating the adiabatic trajectory step by step.

This method was successfully demonstrated by Hegade et al.~\cite{hegade_digitized_2021}, where the authors implemented a digitized adiabatic factorization algorithm on superconducting hardware. They successfully factorized numbers $N=21$, $N=91$, and $N=217$, and also proved enhancements by introducing an additional layer in the algorithm due to application of counter-diabatic driving theories.

\subsubsection{QAOA for Standard Adiabatic Factorization}
In some sense, QAOA can be viewed as a shortcut to adiabaticity implemented in the circuit model for quantum computation. Rather than performing a slow, continuous evolution, QAOA uses a fixed-depth quantum circuit composed of alternating unitaries derived from the mixing and problem Hamiltonians. The parameters of these unitaries are optimized variationally to prepare a state that approximates the solution.

Díez-Valle et al.~\cite{diez-valle_universal_2025} demonstrate that QAOA and quantum annealing share fundamental structural features. Their results provide strong evidence that smooth annealing paths can be reconstructed from the optimal parameters of QAOA, reinforcing the view of QAOA as a digitized, variationally optimized form of adiabatic evolution.

In summary, the ingredients to build and execute the QAOA algorithm applied to integer factorization --- or what we will call the ``\texttt{standard} protocol'' --- are:
\begin{itemize}
    \item A Hamiltonian $\hat{H}_\mathrm{QP}$, Eq.~\eqref{eq:quadratic_problem_hamiltonian}, that encodes the solution to the factorization problem in its ground state.
    \item A mixing Hamiltonian $\hat{H}_\mathrm{M}$ that does not commute with $\hat{H}_\mathrm{QP}$, e.g., Eq.~\eqref{eq:mixing_hamiltonian}.
    \item A quantum circuit that implements the state evolution given by Eq.~\eqref{eq:qaoa_state_evolution}.
    \item A classical optimizer to find the circuit's optimal parameters.
    \item A cost function given by $F_p (\bm{\gamma}, \bm{\beta}) = \bra{\psi_p (\bm{\gamma}, \bm{\beta})} \hat{H}_\mathrm{QP} \ket{\psi_p (\bm{\gamma}, \bm{\beta})}$.
\end{itemize}

\begin{figure}[h]
    \centering
    \includegraphics[width=1\textwidth]{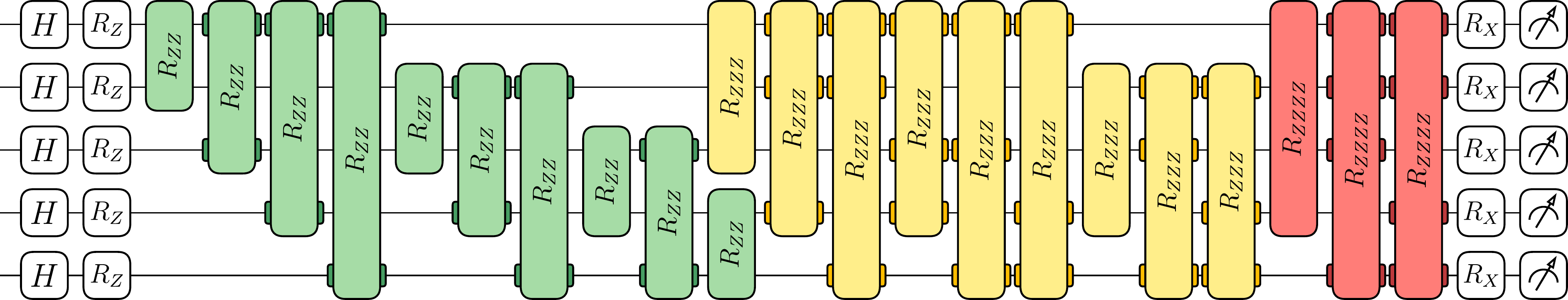}
    \caption{One-layer circuit for factorizing the number $N=35$ using the standard QAOA protocol. Notice the presence
    of three- and four-qubit gates, highlighted in yellow and red, respectively. Rotation angles are omitted for simplicity.}
    \label{fig:standard_circuit}
\end{figure}

Due to the form of Eq.~\eqref{eq:quadratic_problem_hamiltonian}, the problem Hamiltonian can be expressed as:
\begin{equation}
    \hat{H}_\mathrm{QP} = n \1 + \sum_i a_i \hat{\sigma}^z_i
    + \sum_{ij} b_{ij} \hat{\sigma}^z_i \hat{\sigma}^z_j
    + \sum_{ijk} c_{ijk} \hat{\sigma}^z_i \hat{\sigma}^z_j \hat{\sigma}^z_k
    + \sum_{ijk\ell} d_{ijk\ell} \hat{\sigma}^z_i \hat{\sigma}^z_j \hat{\sigma}^z_k \hat{\sigma}^z_\ell
    \label{eq:expanded_problem_quadratic_hamiltonian}
\end{equation}
Then, the part of $e^{-i \gamma \hat{H}_\mathrm{QP}}$ corresponding to the term $a_i \hat{\sigma}^z_i$ is
\begin{equation}
    e^{-i \gamma a_i \hat{\sigma}^z_i} = e^{-i \gamma a_i} \ket{0}\bra{0}
    + e^{i \gamma a_i} \ket{1}\bra{1}
    = R_Z(2 \gamma a_i)\,,
\end{equation}
the part corresponding to $b_{ij} \hat{\sigma}^z_i \hat{\sigma}^z_j$ is
\begin{equation}
    \begin{split}
        e^{-i \gamma b_{ij} \hat{\sigma}^z_i \hat{\sigma}^z_j} \\
         &=e^{-i\gamma b_{ij}} \ket{00}\bra{00} + e^{i\gamma b_{ij}} \ket{10}\bra{10}
        +  e^{i\gamma b_{ij}} \ket{01}\bra{01} + e^{-i\gamma b_{ij}} \ket{11}\bra{11} \\
        &= R_{ZZ} (2 \gamma b_{ij})\,,
    \end{split}
\end{equation}
and so on for the higher-order terms. Similarly, the mixing Hamiltonian $\hat{H}_\mathrm{M}$ gives rise to individual X-rotations at the end of each QAOA layer. With all this, one constructs the QAOA circuit --- such as the circuit shown in Fig.~\ref{fig:standard_circuit} used for factorization of the number $N=35$.

A related study by Anschuetz et al.~\cite{anschuetz_variational_2018} applies a similar method, combined with additional preprocessing and simplifications, to factorize numbers as large as $291311$. Building on this idea, Karamlou et al.~\cite{karamlou_analyzing_2021} used classical preprocessing heuristics to factorize $1099551473989$, $3127$, and $6557$ using only $3$, $4$, and $5$ qubits, respectively. In contrast, our goal is not to optimize for specific instances, but rather to analyze the general, standard approach to quantum factorization without relying on ad-hoc techniques. The motivatioin comes from establishing a broadly applicable framework, valid for any input $N$, while avoiding the need for modifications to the underlying factorization algorithm.

\subsection{QAOA for Linearized Adiabatic Factorization}
\begin{figure}[h]
    \centering
    \includegraphics[width=1\textwidth]{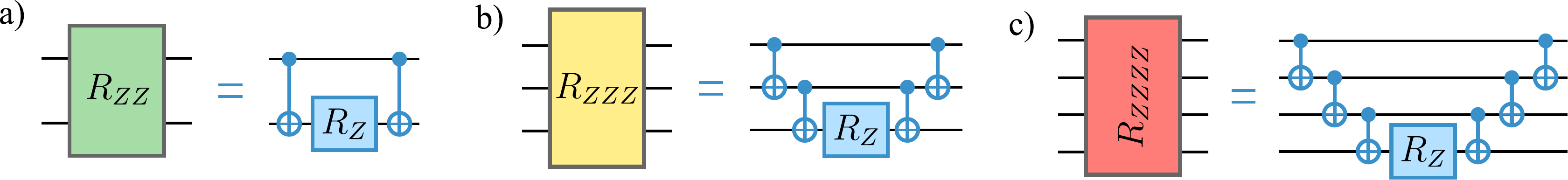}
    \caption{Decomposition of (a) two-, (b) three-, and (c) four-qubit Z-rotation gates in CNOTs and
    single-qubit Z-rotations.}
    \label{fig:gate_decomposition}
\end{figure}

As mentioned in the introduction to Section~\ref{Section:AFA}, the adiabatic factorization algorithm
needs to deal with the difficulty of implementing three- and four-body interaction terms. In the 
digitized version of this problem, like QAOA, this difficulty is bypassed by assuming that all three- and four- qubit gates --- see e.g. Fig.~\ref{fig:standard_circuit} --- can be efficiently implemented in a digital computer. However, for universal computers available so far, these gates are not efficiently implemented and we need to decompose them in multiple two-qubit gates, as shown in Fig.~\ref{fig:gate_decomposition}, what might result in lower fidelities at the end of the quantum circuit in a real scenario.

\begin{figure}[h]
    \centering
    \includegraphics[width=0.45\textwidth]{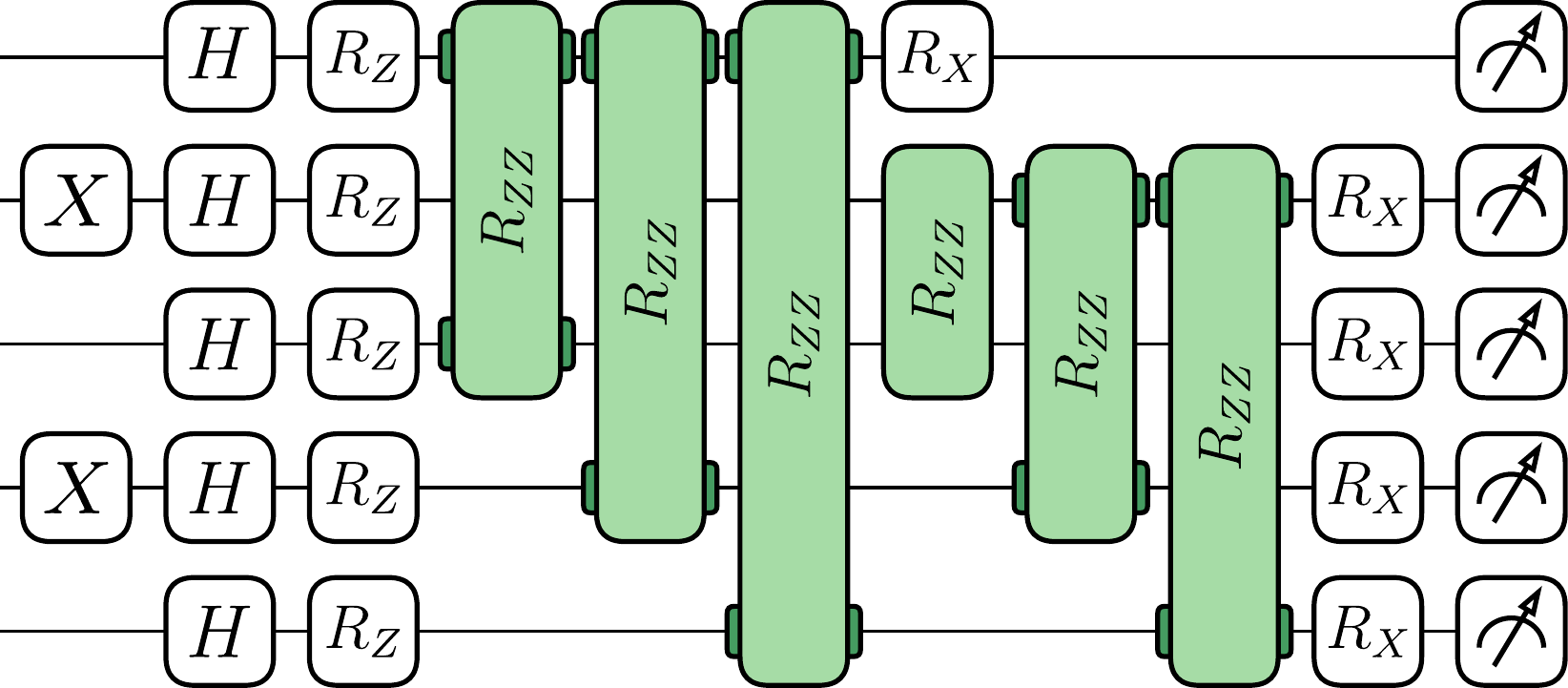}
    \caption{One-layer circuit for factorizing the number $N=35$ using our protocol, evolving the state with the linear Hamiltonian $H_{\mathrm{LP}}$. Notice the simplification with respect to Fig.~\ref{fig:standard_circuit}
    due to the absence of three- and four-qubit gates.}
    \label{fig:linear_circuit_35.pdf}
\end{figure}

To mitigate this issue, we propose a linearized problem Hamiltonian inspired by the same factorization condition, defined as:
\begin{equation}
	\hat{H}_\mathrm{LP} = N \1 - \bigg( \sum_{\ell=1}^{n_p} 2^\ell \hat{x}_\ell + \1 \bigg)
	\bigg( \sum_{m=1}^{n_q} 2^m \hat{y}_m + \1 \bigg)\,.
	\label{eq:linear_problem_hamiltonian}
\end{equation}
Unlike the original Hamiltonian $\hat{H}_\mathrm{QP}$, whose ground state encodes the solution, $\hat{H}_\mathrm{LP}$ contains only two-body terms and is therefore easier to implement (see Fig.~\ref{fig:linear_circuit_35.pdf}), but the factorization solution corresponds to an eigenstate with eigenvalue zero rather than the ground state. In other words, here we consider that the solution of the factorization is encoded in the \textit{null-space} of the Hamiltonian, composed by eigenstates with eigenvalue $0$. This approach does not impose limitations on the algorithm, since the adiabatic theorem is not restricted to ground states but applies to any non-degenerate eigenstate. Therefore, the adiabatic theorem supports that it is possible to target this eigenstate through a suitable adiabatic or variational process. Based on this idea, our proposal is to replace the original Hamiltonian in the QAOA layers with $\hat{H}_\mathrm{LP}$, leveraging the possibility of starting from an eigenstate near the middle of the initial Hamiltonian’s spectrum and evolving under $\hat{H}_\mathrm{LP}$ to preserve this eigenstate structure, ultimately reaching the correct factorization solution, which also lies near the center of the spectrum.

\section{Algorithms Implementation: Methods}
\label{Section:Methods}

In this work, we use the basic formulation of QAOA as a reference framework. To enable a fair comparison between the standard factorization protocol and our alternative variant, we configure the algorithm --- including choices for parameter initialization and classical optimization --- so that the standard protocol performs as effectively as possible. This approach ensures that any observed differences in performance arise from the protocol itself rather than from specific details of the QAOA setup.

For the numerical simulations, several software frameworks are available, including PennyLane and Qiskit. However, in this study we opted for a direct state-vector simulation implemented in NumPy, where circuit layers are applied through explicit matrix multiplications. This strategy avoids the overhead associated with gate-by-gate simulation and provides a more transparent view of the algorithmic dynamics, while remaining computationally efficient at the scale considered here.

To facilitate reproducibility, the complete implementation together with the numerical procedures and data analysis scripts has been made publicly available in a GitHub repository\footnote{\url{https://github.com/fpllcr/enhanced-digitized-adiabatic-quantum-factorization}}.

\subsection{Initial State}
\label{Section:InitialState}

In line with the adiabatic theorem, which states that a quantum system prepared in an eigenstate of a slowly varying Hamiltonian will remain in its instantaneous eigenstate, we aim to initialize the system in an eigenstate of the mixing Hamiltonian $\hat{H}_\mathrm{M}$. Ideally, this initial state should occupy a similar region of the spectrum as the target solution under the problem Hamiltonian $\hat{H}_\mathrm{P}$, so that the variational evolution requires minimal spectral “rearrangement.” The specific choice of initial state depends on the protocol used:

\begin{itemize}
    \item \textbf{\texttt{standard} protocol} ($\hat{H}_\mathrm{P} = \hat{H}_\mathrm{QP}$): The solution corresponds to the ground state of the problem Hamiltonian. To reflect this structure, we initialize the system in the ground state of the mixing Hamiltonian, $$\ket{\psi (0)} = \ket{+}^{\otimes n},$$ which is both easy to prepare and spectrally aligned with low-energy states of $\hat{H}_\mathrm{P}$.

    \item \textbf{\texttt{linear} protocols} ($\hat{H}_\mathrm{P} = \hat{H}_\mathrm{LP}$): Here, the correct solution lies near the center of the spectrum rather than at its bottom. Then, we choose an eigenstate of $\hat{H}_\mathrm{M}$ that is approximately centered in its spectrum, $$\ket{\psi (0)} = \ket{+-+-+\cdots}.$$ For problems with an odd number of qubits, this configuration cannot be perfectly centered, but QAOA's non-adiabatic nature tolerates this slight asymmetry without degrading performance.
\end{itemize}

\subsection{Parameter Initialization}
\label{Section:IncrementalApproach}

In multi-layer QAOA, increasing the number of layers also increases the number of parameters to be optimized. Finding global optima is considerably simpler for $p=1$ (i.e., two parameters) than for circuits with many layers. Moreover, when a new layer is added, the previously optimized solution remains valid within the enlarged parameter space: the parameters corresponding to the first $p$ layers can be initialized with their previously optimized values, while the newly introduced parameters can be set to zero. Consequently, adding an additional layer ($p \rightarrow p+1$) can only maintain or improve performance relative to the previous iteration.

\begin{figure}[h]
    \centering
    \includegraphics[width=0.91\textwidth]{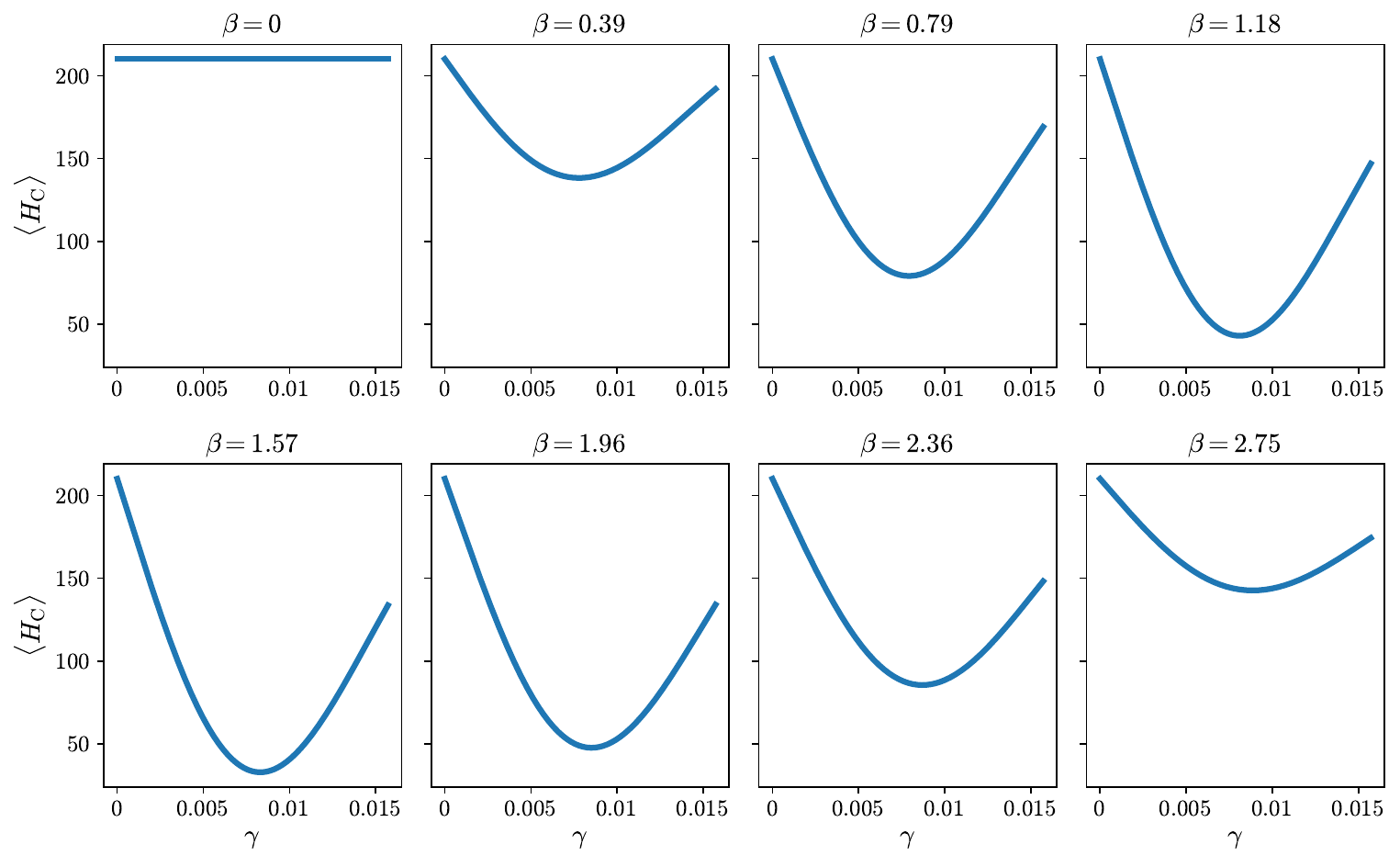}
    \caption{Cost function profile for the quadratic Hamiltonian when $N=21$. After this
    analysis, $\gamma_0=0.0075$ and $\beta_0=1.57$ are used as initial parameters for $p=1$.
    Maximum $\gamma$ is determined as the value that causes angle folding for the
    maximum energy eigenvalue, i.e. $e^{-i \gamma_\mathrm{max} E_\mathrm{max}} = e^{-2\pi i}$.}
    \label{fig:costfun_landscape}
\end{figure}

This observation motivates a progressive training strategy, often referred to as an \emph{incremental} approach, which proceeds as follows~\cite{zhou_quantum_2020}:
\begin{enumerate}
    \item Analyze the cost function landscape for $p=1$ to select initial parameters $\gamma_1$ and $\beta_1$ close to the optimal region (Fig.~\ref{fig:costfun_landscape}).
    \item Optimize QAOA with $p=1$ to obtain the parameters $\widetilde{\gamma}_1$ and $\widetilde{\beta}_1$.
    \item Optimize QAOA with $p=2$, initializing $\gamma_1=\widetilde{\gamma}_1$ and $\beta_1=\widetilde{\beta}_1$, and selecting suitable initial values for $\gamma_2$ and $\beta_2$ using an appropriate heuristic. This yields the updated parameters $\widetilde{\gamma}_1, \widetilde{\gamma}_2, \widetilde{\beta}_1, \widetilde{\beta}_2$.
    \item Repeat this procedure iteratively up to the desired depth $p$, each time initializing
    with the optimized parameters from the previous iteration and applying a strategy to set
    the new layer's parameters.
\end{enumerate}

As aforementioned, adding a new QAOA layer requires a suitable heuristic to initialize the additional variational parameters. Beyond random initialization, two specific strategies have been considered in this work:

\begin{itemize}
    \item \textbf{Parameter interpolation.}  
    This method, adopted by Zhou et al.~\cite{zhou_quantum_2020}, linearly interpolates the trajectory traced by the optimized parameters at depth $p$ to generate an initial guess for depth $p+1$. Such interpolation is particularly effective for problems like MaxCut and QUBO, where the optimal parameters often evolve monotonically in an adiabatic-like fashion.

    \item \textbf{Alternative heuristic.}  
    This approach, which achieved superior performance for both the standard and linearized protocols in our experiments, initializes the new parameters as
    \begin{equation}
        \begin{cases}
            \gamma_{p+1} \leftarrow \widetilde{\gamma}_{p}, \\
            \beta_{p+1} \leftarrow 0.
        \end{cases}
        \label{eq:initialization_heuristic}
    \end{equation}
\end{itemize}

Two factors motivated the adoption of this alternative strategy.  
First, the interpolation heuristic performs poorly for the linearized protocol, frequently leading to trapping in local minima.  
Second, integer factorization differs fundamentally from MaxCut and QUBO problems: there is no evidence that optimal parameters follow adiabatic-like trajectories, and our results confirm that they do not (see Fig.~\ref{fig:optimizer_comparison} and Fig.~\ref{fig:optimizer_parameter_comparison}).  
The fact that the alternative heuristic also outperforms interpolation for the standard protocol ensures a fair basis for comparison, as both protocols are evaluated under the most favorable optimization setup for the reference (standard) implementation.

\subsection{Classical optimizer}

A wide variety of classical optimizers have been employed within QAOA, as reviewed in the literature~\cite{blekos_review_2024}.  
In this work, we focused on optimizers available in the SciPy library and tested representative classes: bounded vs.\ unbounded and gradient-free vs.\ gradient-based.  

\begin{figure}[h]
    \centering
    \includegraphics[width=0.99\textwidth]{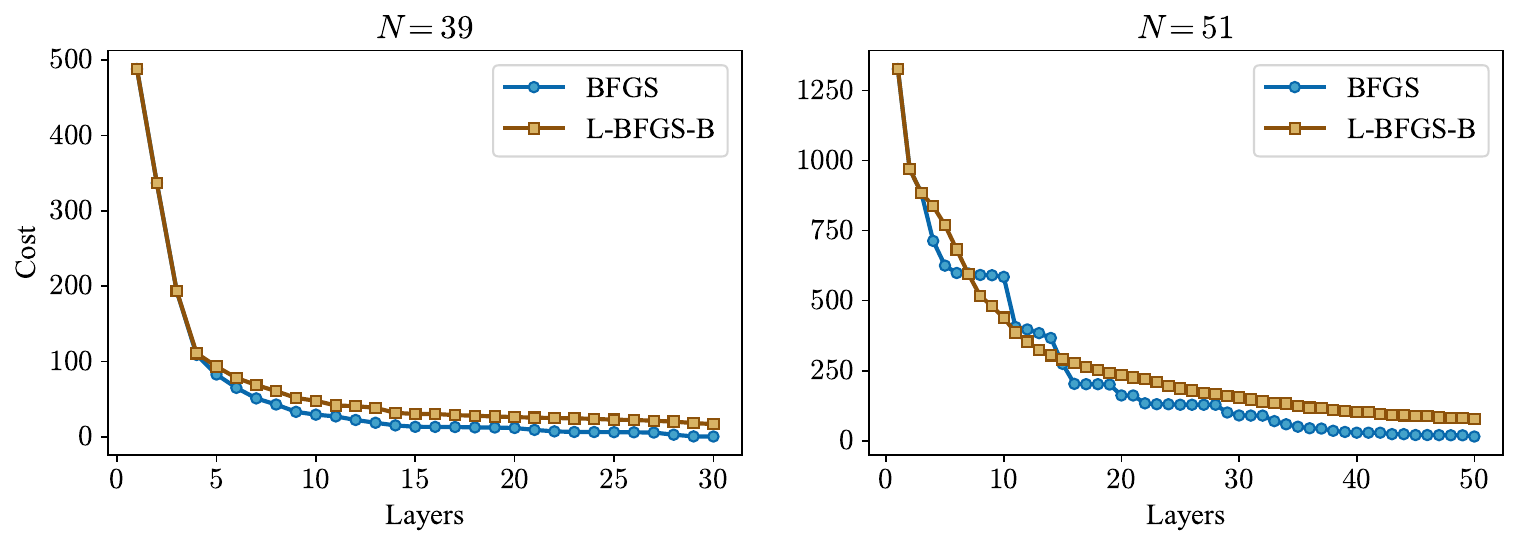}
    \caption{Cost evolution comparison between BFGS and L-BFGS-B for factorizing $N=39$ and $N=51$ using the standard protocol. The performance gap was observed consistently up to 8-qubit problems.}
    \label{fig:optimizer_comparison}
    \includegraphics[width=1\textwidth]{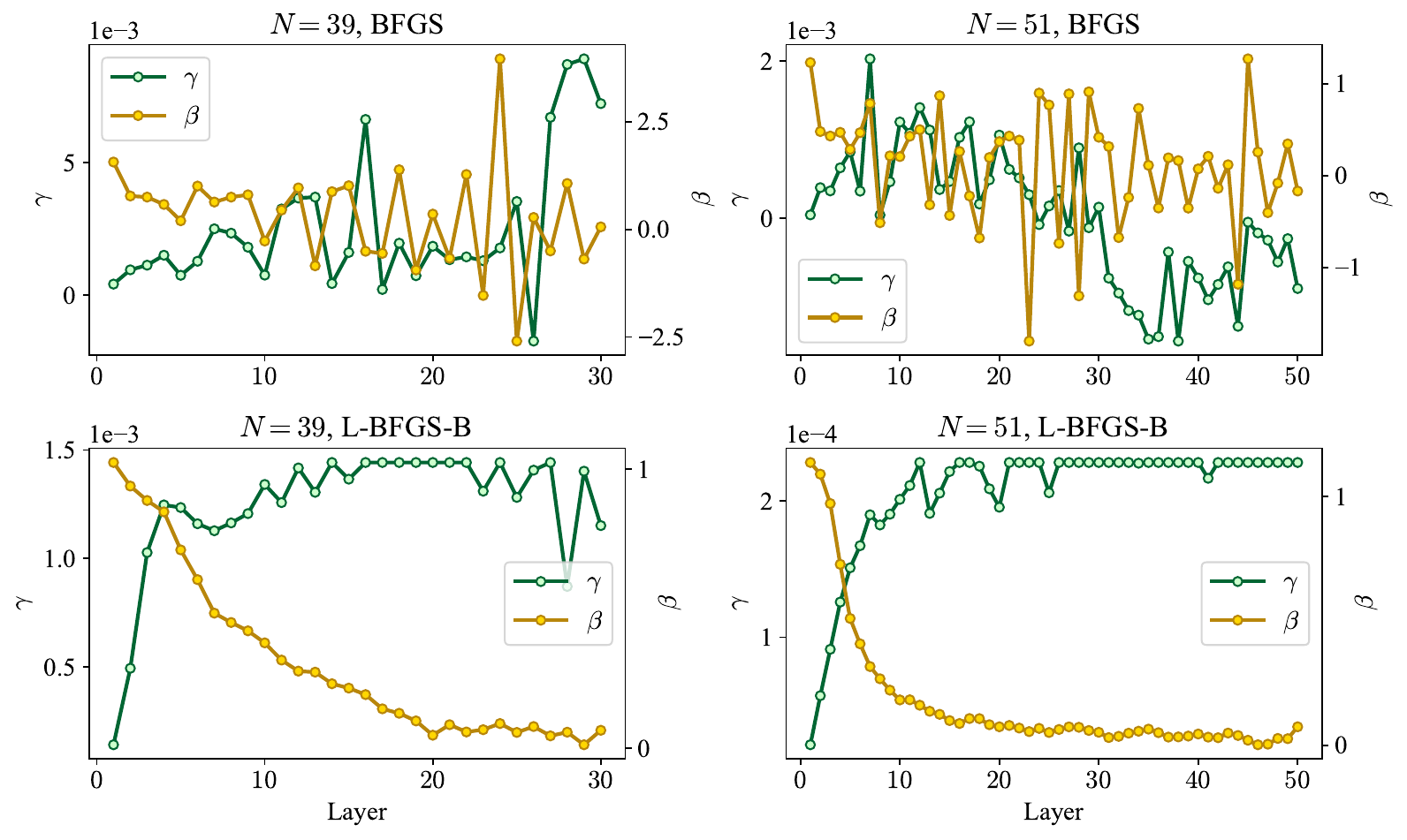}
    \caption{Variational parameter evolution for $N=39$ and $N=51$ using BFGS and L-BFGS-B. An adiabatic-like trajectory emerges under L-BFGS-B due to parameter bounds.}
    \label{fig:optimizer_parameter_comparison}
\end{figure}

Gradient-free methods such as Nelder-Mead showed promising performance for small problems but scale poorly with the number of parameters, and were therefore discarded. Cobyla exhibited poor convergence and frequent trapping in local minima, even for few-qubit cases, making it unsuitable for this study. Among gradient-based methods, the best performance was obtained with BFGS and its bounded variant L-BFGS-B:

\begin{itemize}
    \item \textbf{L-BFGS-B.}  
    This bounded optimizer is well-suited for QUBO and MaxCut problems, where parameter symmetries justify restricting $\gamma$ and $\beta$ to specific intervals. Under such constraints, L-BFGS-B often produces adiabatic-like parameter trajectories, with $\gamma$ starting near zero and increasing smoothly, while $\beta$ decreases monotonically toward zero.  

    \item \textbf{BFGS.}  
    As an unbounded optimizer, BFGS allows variational parameters to take any real value. In all tested cases up to 8~qubits, BFGS consistently outperformed L-BFGS-B, likely because its unbounded nature helps it escape local minima that trap the bounded optimizer (see Fig.~\ref{fig:optimizer_comparison}).
\end{itemize}

Based on these findings, BFGS is selected as the classical optimizer for all experiments.  
This choice complements the other elements of our QAOA setup, ensuring a consistent and high-performance framework for comparing the standard protocol and the proposed linearized protocol.

In summary, the methodological strategy to implement all QAOA circuits adopted through this thesis is as follows:
\begin{itemize}
    \item \textbf{Initial state:} $\ket{+}^{\otimes n}$ for the \texttt{standard} protocol, and $\ket{+-+-+\cdots}$ for \texttt{linear} protocol. 
    \item \textbf{Layer-by-layer training:} QAOA parameters are optimized incrementally, starting from $p=1$ and using previously optimized parameters as initialization for deeper circuits.  
    \item \textbf{Parameter initialization:} New layers are initialized using the heuristic $\gamma_{p+1} \leftarrow \widetilde{\gamma}_p$, $\beta_{p+1} \leftarrow 0$, which outperforms interpolation strategies.  
    \item \textbf{Classical optimizer:} The gradient-based, unbounded BFGS method is employed, providing superior convergence compared to bounded alternatives.  
\end{itemize}

Together, these decisions define a robust and fair QAOA implementation, enabling meaningful comparisons between different problem Hamiltonians and protocols.

\chapter{Results}
\label{Chapter:Results}

\dropcap{T}his chapter presents the numerical experiments performed to compare the performance
of the standard QAOA protocol with the proposed linearized protocol. The standard protocol is
evaluated in its original formulation, while the linearized protocol is analyzed under two different
cost functions to determine which provides the most favorable performance. Table~\ref{tab:protocols_overview}
summarizes the three protocols considered throughout this chapter. 

\begin{table}[h]
    \centering
    \begin{tabular}{@{}cccc@{}}
        \toprule
        Protocol            & Mixing Hamiltonian    & Problem Hamiltonian   & Cost function \\
        \midrule
        \texttt{standard}            & $\hat{H}_\mathrm{M}$   & $\hat{H}_\mathrm{QP}$ & $\langle \hat{H}_\mathrm{QP} \rangle$ \\
        \texttt{linear\_quadratic}   & $\hat{H}_\mathrm{M}$   & $\hat{H}_\mathrm{LP}$ & $\langle \hat{H}_\mathrm{QP} \rangle$ \\
        \texttt{linear\_abs}         & $\hat{H}_\mathrm{M}$   & $\hat{H}_\mathrm{LP}$ & $\langle |\hat{H}_\mathrm{LP}| \rangle$ \\
        \bottomrule
    \end{tabular}
    \caption{Overview of QAOA protocols analyzed in this work.}
    \label{tab:protocols_overview}
\end{table}

The primary evaluation metric throughout this chapter is the \emph{fidelity}, which directly
corresponds to the probability of successfully obtaining the correct factors of the integer
to be factorized. Using fidelity as a baseline metric allows for a clear and consistent
comparison between protocols.

All practical aspects of the QAOA setup --- including initialization heuristics and the choice of
classical optimizer --- were fixed and justified in section~\ref{Section:Methods} to ensure
fair and consistent comparisons. The focus here is exclusively on evaluating and contrasting
the protocols across problem instances of increasing size under these controlled conditions.

As representative examples, we present detailed results for the semiprime integers $N = 25$, $N = 77$,
and $N = 143$. A complete set of results can be found in the accompanying GitHub repository
and is summarized in the appendix~\ref{Chapter:Appendix}. Table~\ref{tab:instances_overview} provides an overview of
these selected test cases, including relevant problem parameters and solution encodings.
\begin{table}[h]
    \centering
    \begin{tabular}{@{}ccccccccc@{}}
        \toprule
        $N$ & $n$ (\# qubits) & $p$ ($p'$) & $q$ ($q'$) & $n_p$ & $n_q$ & $p_\mathrm{bitstring}$ & $q_\mathrm{bitstring}$ & solution(s) \\
        \midrule
        25  & 4          & 5 (2)      & 5 (2)      & 2     & 2     & 10                     & 10                     & 0101                  \\
        77  & 6          & 7 (3)      & 11 (5)     & 2     & 4     & 11                     & 0101                   & 111010                \\
        143 & 8          & 11 (5)     & 13 (6)     & 3     & 5     & 101                    & 00110                  & 10101100, 01110100    \\
        \bottomrule
    \end{tabular}
    \caption{Overview of semiprime test instances used as representative examples.}
    \label{tab:instances_overview}
\end{table}

\begin{figure}[h]
    \centering
    \includegraphics[width=1\textwidth]{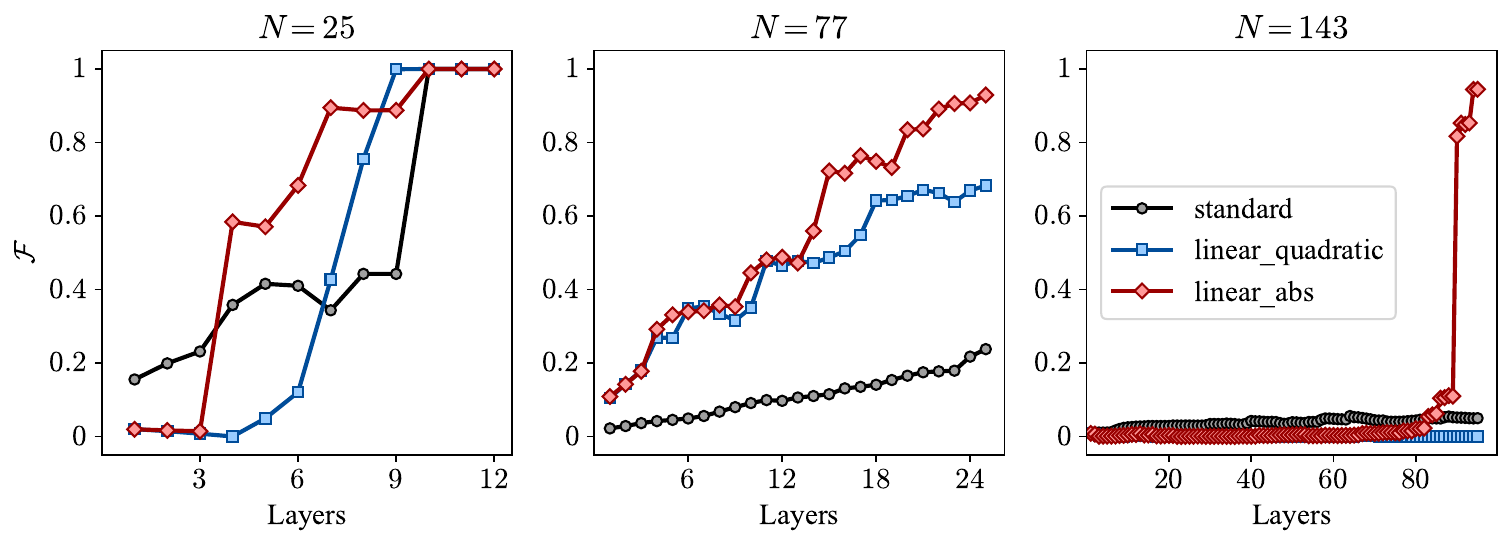}
    \caption{Fidelity versus QAOA layer depth. Larger problems require deeper circuits to achieve significant fidelities.}
    \label{fig:fidelity_layers}
\end{figure}

Firstly, we evaluate the fidelities achieved by the different protocols as a function of the number of QAOA layers (Fig.~\ref{fig:fidelity_layers}). Three distinct trends are observed:  
\begin{itemize}
    \item For $N = 25$, the \texttt{standard} protocol performs better at shallow depths, but both \texttt{linear} protocols eventually surpass it.  
    \item For $N = 77$, both \texttt{linear} protocols consistently outperform the \texttt{standard} protocol across all layers.  
    \item For $N = 143$, the \texttt{standard} protocol initially leads, until a sudden fidelity jump in one of the \texttt{linear} protocols overtakes it.  
\end{itemize}

As the problem size increases, the third behavior---a sudden jump in fidelity---appears more frequently (see Fig.~\ref{fig:fidelity_layers_all}). Although the cost-function landscape evolves more smoothly (see Fig.~\ref{fig:cost_layers_all}), the fidelity of \texttt{linear} protocols often exhibits these abrupt improvements. The underlying reasons for this phenomenon will be analyzed in Chapter~\ref{Chapter:Discussion}.

\begin{figure}[h]
    \centering
    \includegraphics[width=1\textwidth]{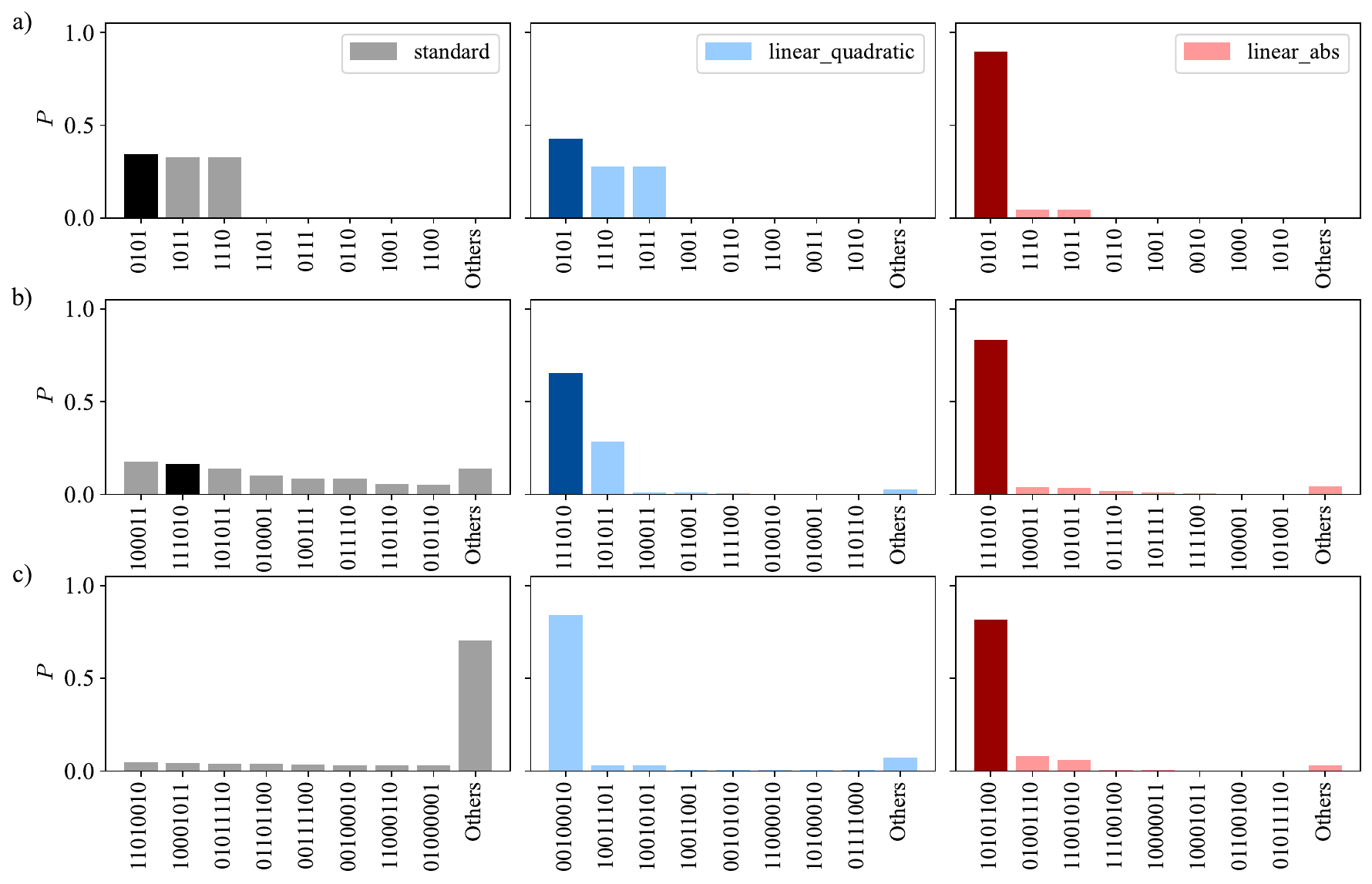}
    \caption{Final state populations for (a) $N=25$, (b) $N=77$, and (c) $N=143$. The number
    of layers was fixed as the minimum required to reach at least 80\% fidelity by any
    protocol. Solution states are shown as dark-colored bars.}
    \label{fig:populations}
\end{figure}

To complement the fidelity analysis, Fig.~\ref{fig:populations} depicts the state populations at the end of each protocol, with valid solutions highlighted by dark-colored bars.

\begin{figure}[h]
    \centering
    \includegraphics[width=1\textwidth]{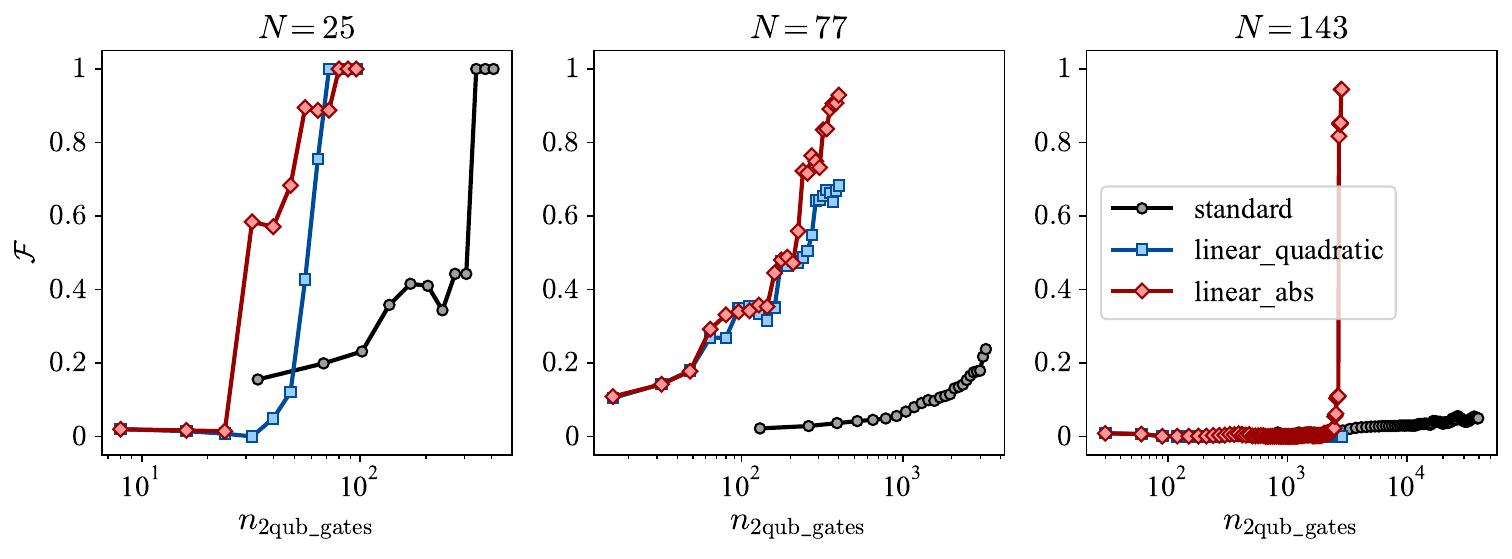}
    \caption{Fidelity versus the number of two-qubit gates. This metric highlights the resource efficiency of the \texttt{linear} protocols.}
    \label{fig:fidelity_gates}
\end{figure}

Although fidelity as a function of depth already demonstrates the improved performance of the \texttt{linear} protocols, it is essential to evaluate them in terms of quantum resources --- which is the central goal of this work. Specifically, we compare protocols using the number of two-qubit gates as the resource metric (Fig.~\ref{fig:fidelity_gates}). In these terms, the advantage of the \texttt{linear} protocols becomes clearer: they achieve significantly higher fidelities while requiring far fewer quantum operations.  

\chapter{Discussion}
\label{Chapter:Discussion}

The numerical results allow us to assess how the proposed linearized QAOA protocols compare to the standard formulation and to evaluate the role of the cost function in shaping their performance.

\subsection*{Standard versus linear\_quadratic protocols}

Since both the \texttt{standard} and \texttt{linear\_quadratic} protocols are optimized with respect to the same cost function, $\langle \hat{H}_\mathrm{QP} \rangle$, their performance can be directly compared. In most cases, the data indicate that the \texttt{linear\_quadratic} protocol reaches relevant solutions with acceptable fidelity at shallower circuit depths. Moreover, when performance is measured against the number of two-qubit gates --- a more meaningful indicator of quantum resource requirements --- the linear protocol consistently requires fewer resources to achieve comparable or superior fidelities.

However, in some problem instances like $N=143$ in Fig.~\ref{fig:fidelity_layers} and Fig.~\ref{fig:fidelity_gates}, it may initially appear that the algorithm performs better under the \texttt{standard} protocol than under \texttt{linear\_quadratic}. In terms of fidelity alone, this is technically correct, but the fidelities in this regime are so low that the difference is not practically significant. In contrast, the cost plots in Fig.~\ref{fig:cost_layers_all} reveal a different picture: the \texttt{linear\_quadratic} protocol consistently achieves lower cost values. This behavior indicates that it is progressing toward a better population transfer to the target state, which should ultimately translate into higher fidelity at later stages.

\subsection*{Impact of the cost function}

The choice of cost function proves to be a decisive factor. Between the two linearized protocols, \texttt{linear\_abs} generally outperforms \texttt{linear\_quadratic}, achieving higher fidelities at lower resource counts. This highlights that even within a fixed ansatz structure, selecting a cost function that better reflects the problem's energy landscape can significantly enhance performance. This observation could motivate further investigation into whether alternative ways of defining the cost function might offer performance benefits.

\subsection*{Sudden fidelity jumps in linear protocols}

\begin{figure}[h]
    \centering
    \includegraphics[width=0.98\textwidth]{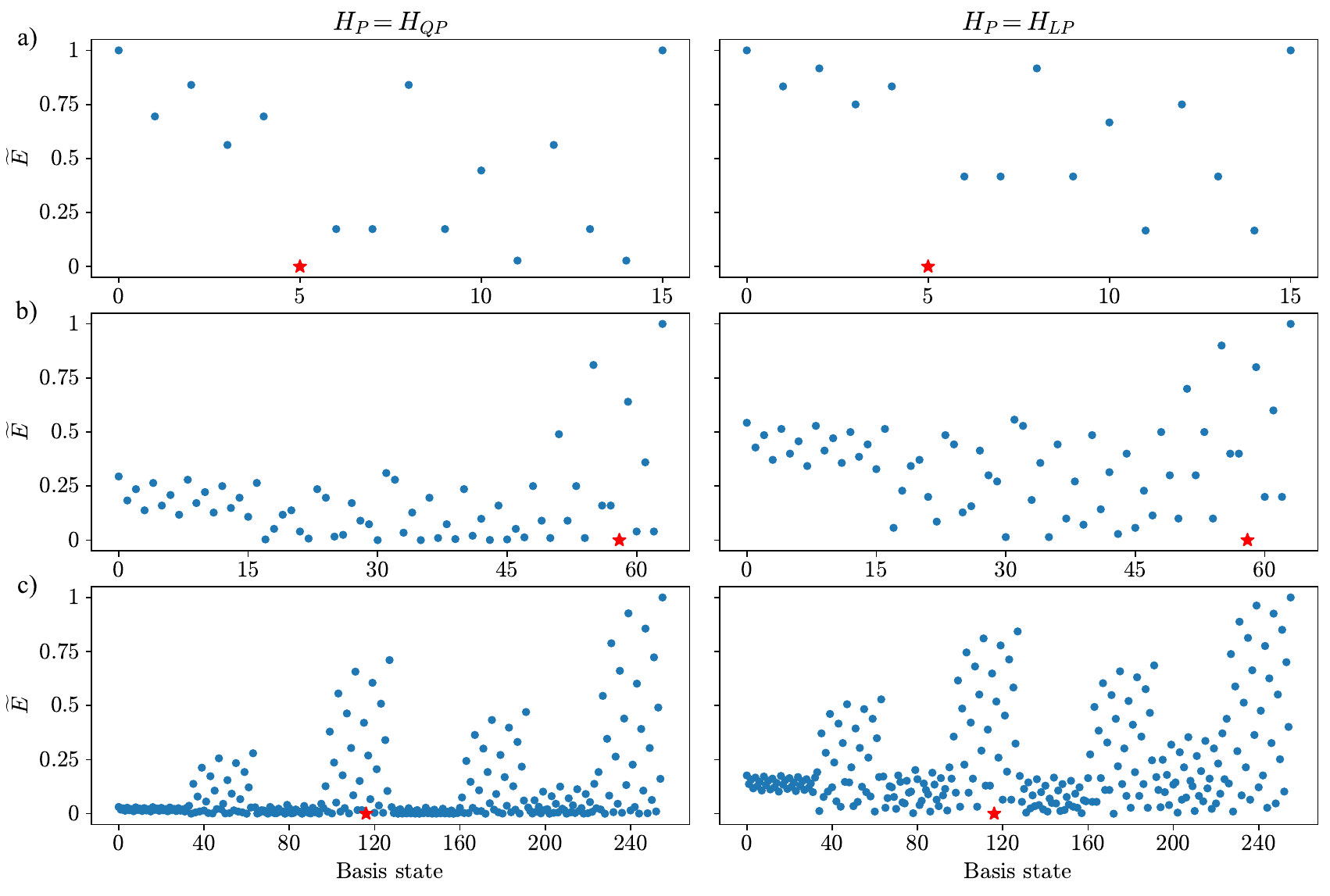}
    \caption{Energy spectrums for (a) $N=25$, (b) $N=77$, and (c) $N=143$. The horizontal axis shows the decimal encoding of computational basis states, while the vertical axis indicates the normalized eigenenergies. The eigenstate corresponding to the solution ($\widetilde{E}=0$) is marked as a red star.}
    \label{fig:spectrums}
\end{figure}

One of the most distinctive features of the linear protocols is the occurrence of abrupt fidelity jumps as the number of QAOA layers increases. This contrasts with the smoother fidelity evolution observed for the standard protocol, a behavior consistently seen across the problem sizes studied (see Fig.~\ref{fig:fidelity_layers_all}). To understand this behavior, we analyzed the spectra of the problem Hamiltonians $\hat{H}_\mathrm{LP}$ and $\hat{H}_\mathrm{QP}$, shown in Fig.~\ref{fig:spectrums}.
We find that the spectrum of $\hat{H}_\mathrm{LP}$ is more widely spread, leading to larger separations between relevant energy levels. Such a structure tends to promote population transfer in discrete steps, which aligns with the observed fidelity jumps. Conversely, the denser spectrum of $\hat{H}_\mathrm{QP}$ supports a more gradual redistribution of amplitude during the variational evolution.

This qualitative picture is further supported by the numerical results summarized in Table~\ref{table:rms}. To quantify the spread of the spectra, we use the root-mean-square (RMS) distance of the eigenvalues with respect to the target energy, which provides a compact measure of how widely distributed the eigenvalues are for each Hamiltonian:
\begin{equation}
    \textrm{RMS} = \sqrt{\dfrac{1}{d} \sum_{i=1}^{d}\widetilde{E}_i^2}\,,
    \label{eq:rms}
\end{equation}
where $\widetilde{E}_i$ is the normalized energy of the $i$-th eigenstate and $d$ the dimension of the Hilbert space asociated to the problem instance.

The results confirm that $\hat{H}_\mathrm{LP}$ consistently exhibits larger RMS distances than $\hat{H}_\mathrm{QP}$ across the problem sizes studied, in agreement with the visual differences in Fig.~\ref{fig:spectrums}. Together, the table and spectra illustrate that the wider separation of eigenvalues in $\hat{H}_\mathrm{LP}$ explains the step-like fidelity behavior observed in the linear protocol.

\begin{table}[h]
    \centering
    \begin{tabular}{@{}lcccccc@{}}
        \toprule
        & \multicolumn{6}{c}{$n$} \\
        \cmidrule(lr){2-7}
        Protocol            & 3     & 4     & 5     & 6     & 7     & 8 \\
        \midrule
        \texttt{standard}   & 0.59  & 0.58  & 0.26  & 0.24  & 0.20  & 0.21 \\
        \texttt{linear}     & 0.70  & 0.68  & 0.41  & 0.37  & 0.32  & 0.32 \\
        \bottomrule
    \end{tabular}
    \caption{Root-mean-square (RMS) distance of the energy spectrum from the target state, used as a measure of how widely the eigenvalues are distributed for each Hamiltonian and problem size.}
    \label{table:rms}
\end{table}

\section{Future prospects}

While the current study focuses on semiprime integers $N$ of moderate size, extending the analysis to larger numbers is pivotal to properly determine whether our results persist in more challenging regimes of factorization. Such investigations would provide a clearer assessment of the asymptotic potential of linearized QAOA and inform the design of more resource-efficient formulations. From an experimental standpoint, our approach offers the additional advantage of requiring only simple quantum circuits, with operations involving at most two qubits. This reduction in circuit complexity significantly enhances the feasibility of near-term realizations on existing hardwares.


\appendix
\chapter{Appendix}
\label{Chapter:Appendix}
\section{Information Table for All Problem Instances}
\label{Section:AppendixInformationTable}

\begin{center}
    \begin{tabular}{@{}ccccccccc@{}}
        \toprule
        $N$ & $n$ (\# qubits) & $p$ ($p'$) & $q$ ($q'$) & $n_p$ & $n_q$ & $p_\mathrm{bitstring}$ & $q_\mathrm{bitstring}$ & solution(s) \\
        \midrule
        15  & 3          & 3 (1)      & 5 (2)      & 1     & 2     & 1                      & 10                     & 101                   \\
        21  & 3          & 3 (1)      & 7 (3)      & 1     & 2     & 1                      & 11                     & 111                   \\
        25  & 4          & 5 (2)      & 5 (2)      & 2     & 2     & 10                     & 10                     & 0101                  \\
        35  & 5          & 5 (2)      & 7 (3)      & 2     & 3     & 10                     & 011                    & 01110, 11010          \\
        39  & 5          & 3 (1)      & 13 (6)     & 2     & 3     & 01                     & 110                    & 10011                 \\
        51  & 6          & 3 (1)      & 17 (8)     & 2     & 4     & 01                     & 1000                   & 100001                \\
        77  & 6          & 7 (3)      & 11 (5)     & 2     & 4     & 11                     & 0101                   & 111010                \\
        87  & 7          & 3 (1)      & 29 (14)    & 3     & 4     & 001                    & 1110                   & 1000111               \\
        95  & 7          & 5 (2)      & 19 (9)     & 3     & 4     & 010                    & 1001                   & 0101001               \\
        115 & 8          & 5 (2)      & 23 (11)    & 3     & 5     & 010                    & 01011                  & 01011010              \\
        119 & 8          & 7 (3)      & 17 (8)     & 3     & 5     & 011                    & 01000                  & 11000010              \\
        143 & 8          & 11 (5)     & 13 (6)     & 3     & 5     & 101                    & 00110                  & 10101100, 01110100    \\
        \bottomrule
    \end{tabular}
    \captionof{table}{Overview of all semiprime test instances studied in this work.}
    \label{tab:instances_overview_all}
\end{center}

\clearpage
\newpage
\section{Fidelity Plots}
\label{Section:FidelityPlots}

\subsection*{Fidelity vs Layers}
\begin{figure}[H]
    \centering
    \includegraphics[width=0.99\textwidth]{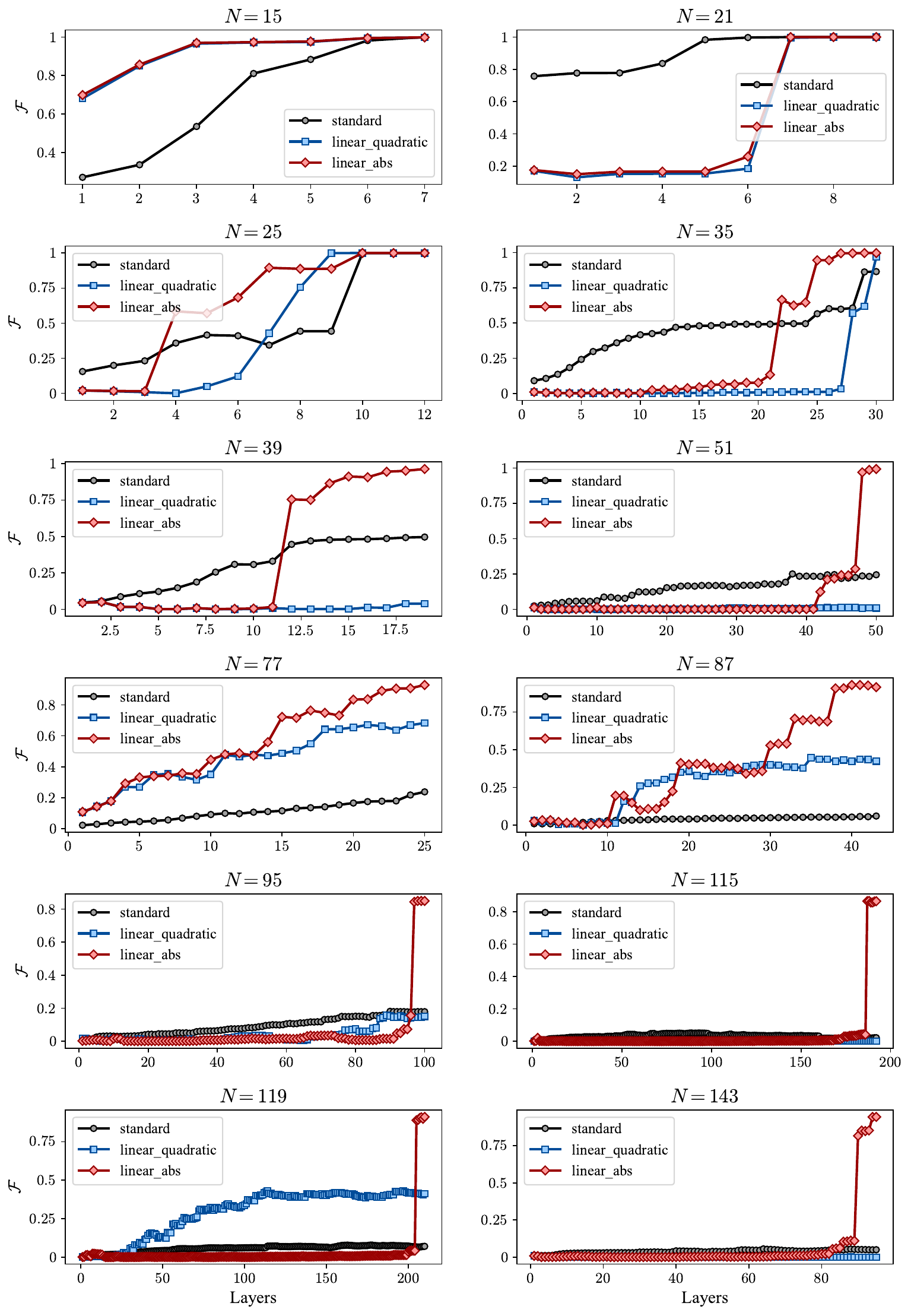}
    \caption{Fidelity versus QAOA layer depth for all factorization instances.}
    \label{fig:fidelity_layers_all}
\end{figure}

\subsection*{Fidelity vs Number of Two-Qubit Gates}
\begin{figure}[H]
    \centering
    \includegraphics[width=0.99\textwidth]{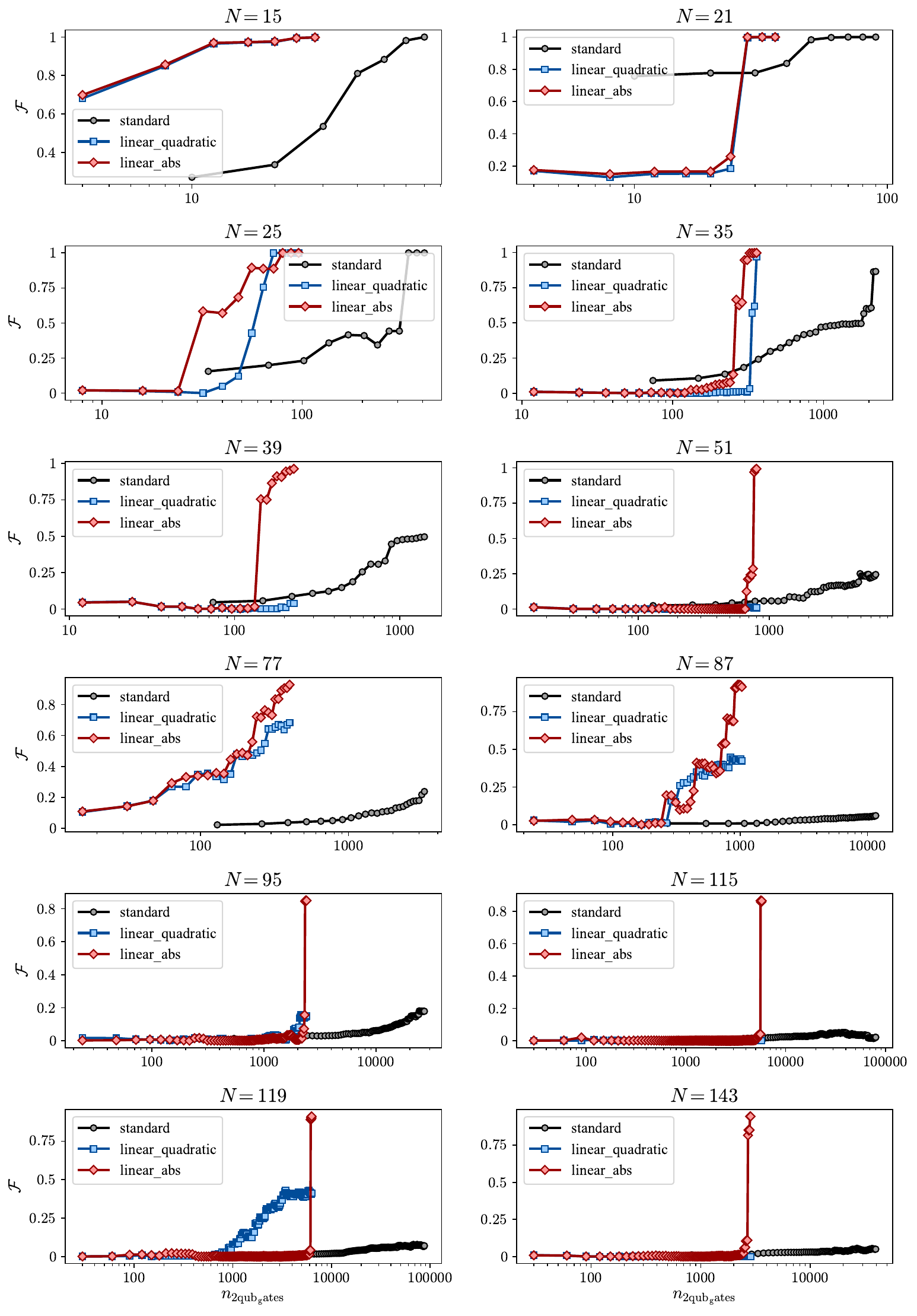}
    \caption{Fidelity versus number of two-qubit gates for all factorization instances.}
    \label{fig:fidelity_gates_all}
\end{figure}
\newpage
\section{Cost vs Layers}
\label{Section:CostPlots}

\begin{figure}[H]
    \centering
    \includegraphics[width=0.99\textwidth]{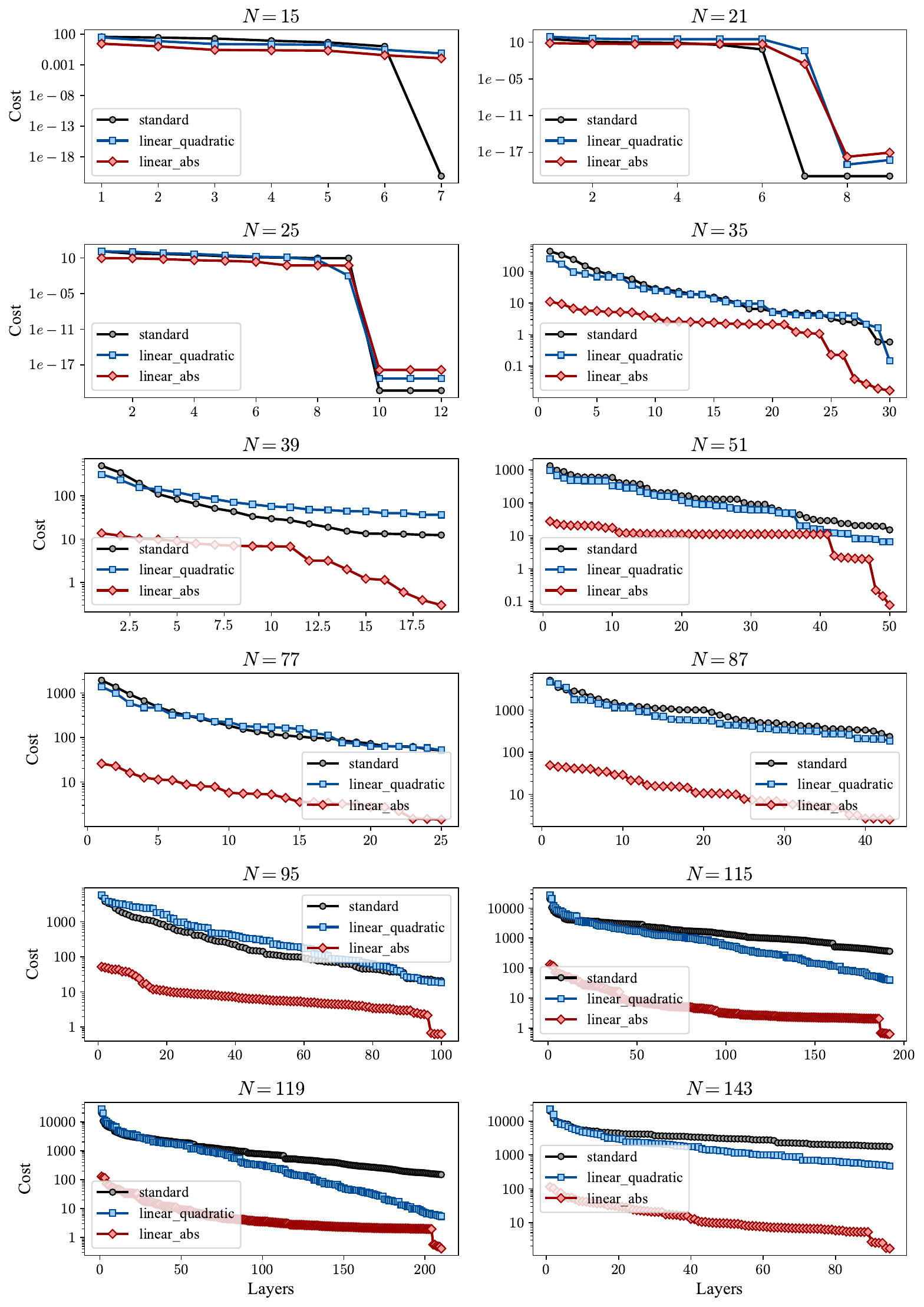}
    \caption{Cost function versus QAOA layer depth for all factorization instances.
    Notice that \texttt{linear\_abs} protocol has a different cost function from
    \texttt{standard} and \texttt{linear\_quadratic}, and therefore they are not
    directly comparable.}
    \label{fig:cost_layers_all}
\end{figure}
\newpage
\section{Populations}
\label{Section:PopulationPlots}

\begin{figure}[H]
    \centering
    \includegraphics[width=1\textwidth]{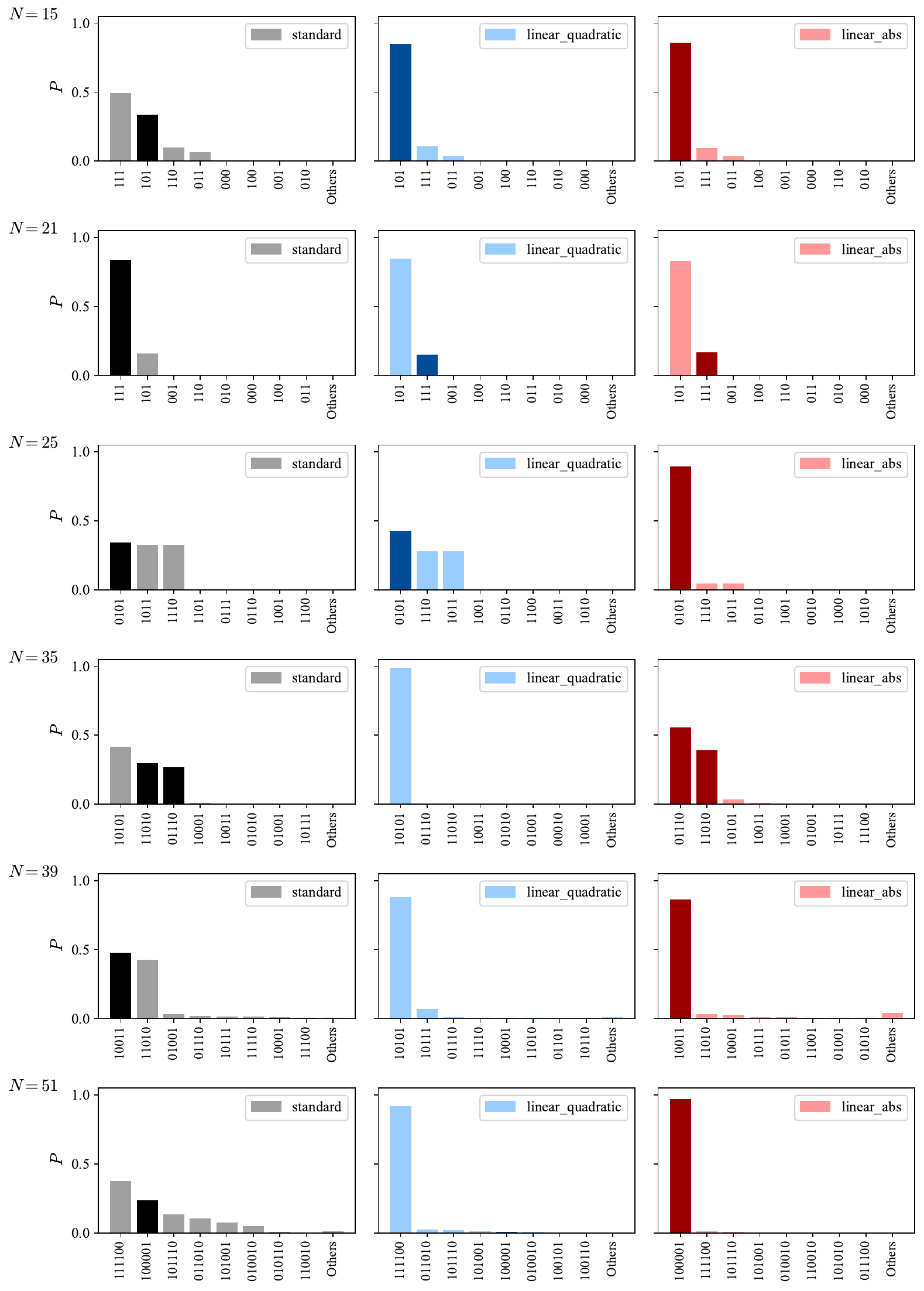}
\end{figure}

\begin{figure}[H]
    \centering
    \includegraphics[width=1\textwidth]{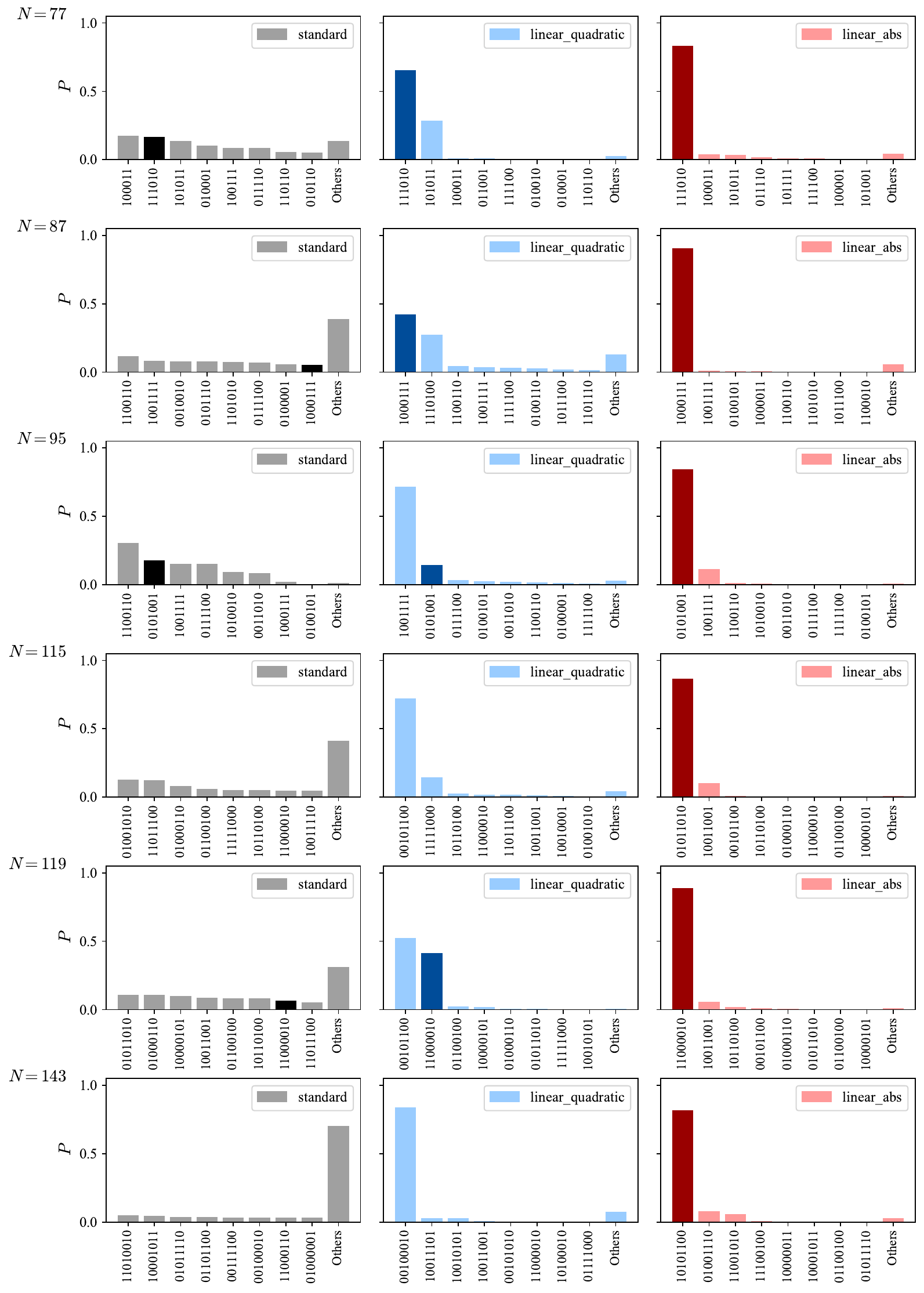}
    \caption{Final state populations for all factorization instances. The number
    of layers was fixed as the minimum required to reach at least 80\% fidelity by any
    protocol. Solution states are shown as dark-colored bars.}
    \label{fig:populations_all}
\end{figure}
\newpage
\section{Energy Spectrums}
\label{Section:EnergySpectrums}

\begin{figure}[H]
    \centering
    \includegraphics[width=0.94\textwidth]{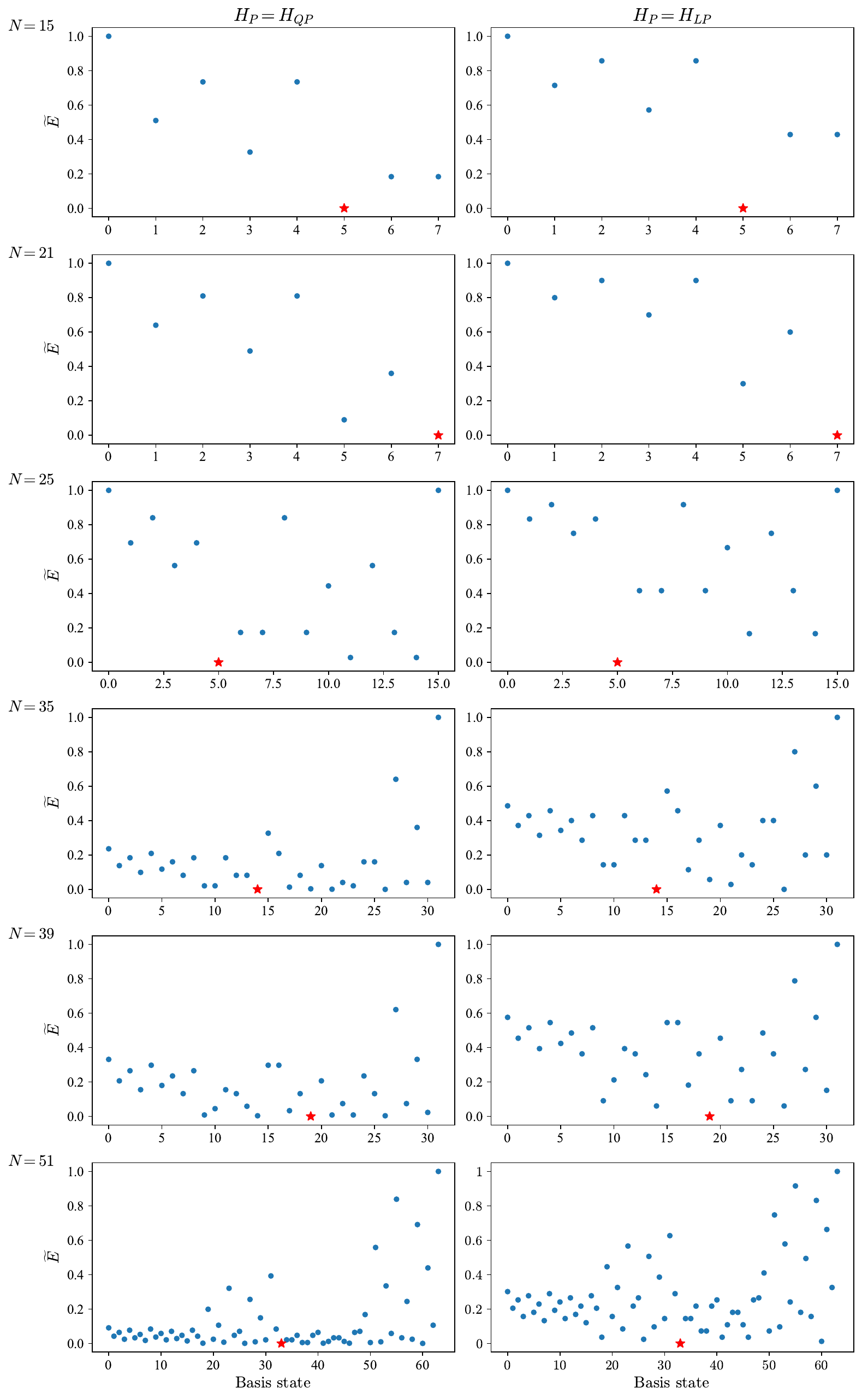}
\end{figure}

\begin{figure}[H]
    \centering
    \includegraphics[width=0.94\textwidth]{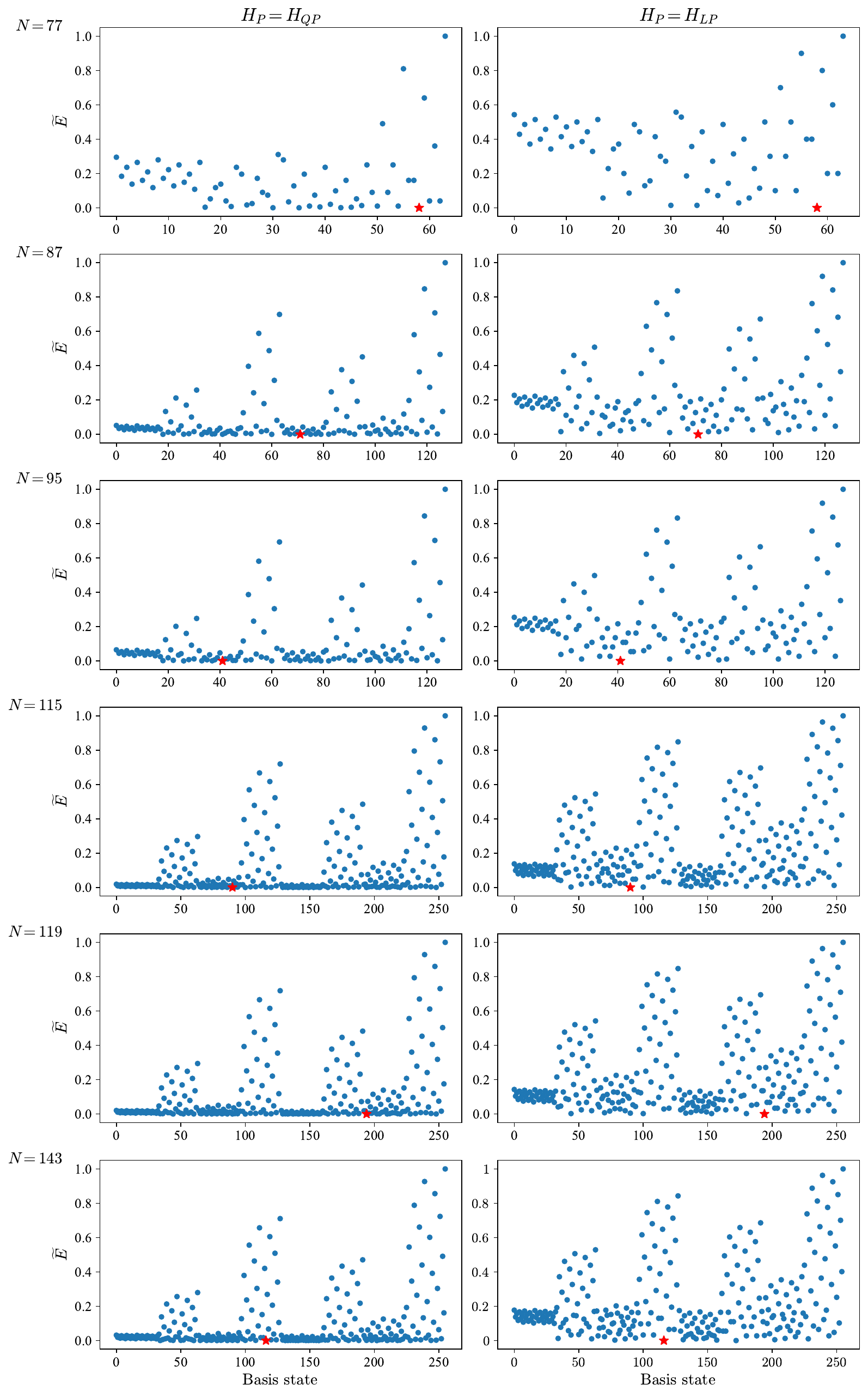}
    \caption{Energy spectrums for all factorization instances. The horizontal axis shows the decimal encoding of computational basis states, while the vertical axis indicates the normalized eigenenergies. Eigenstates corresponding to solutions are marked as red stars.}
    \label{fig:energy_spectrums_all}
\end{figure}

\thumbfalse


\bibliographystyle{unsrt}
\bibliography{bibliography}
\newpage

\end{document}